\documentclass[numbook,natbib,final,runningheads]{svjour3} 
\usepackage{natbib}
\usepackage{epsfig}
\usepackage{graphicx,mathptmx}
\usepackage{color}
\usepackage{url}

\usepackage{makeidx}
\makeindex

\newcommand{\degr}{\ifmmode^\circ\else$^\circ$\fi}

\newcommand{\lapprox} {\, \lower3pt\hbox{$\sim$}\llap{\raise2pt\hbox{$<$}}\,}
\newcommand{\gapprox} {\, \lower3pt\hbox{$\sim$}\llap{\raise2pt\hbox{$>$}}\,}

\include{macros_new}

\begin{document}

\title{Energy Release and Particle Acceleration in Flares:\\
Summary and Future Prospects}

\author{R.~P. Lin$^{1,2}$}

\institute{$^{1}$Physics Department and Space Sciences Laboratory, UC Berkeley, Berkeley, CA 94720\\
$^2$School of Space Research, Kyung Hee University,  Korea\\
\email{rlin@ssl.berkeley.edu} }

\titlerunning{Summary and Future Prospects}
\authorrunning{R.~P. Lin}
\maketitle


\begin{abstract}
\textit{RHESSI} measurements relevant to the fundamental processes of energy release and particle acceleration in flares are summarized. \textit{RHESSI's} precise measurements of hard X-ray continuum spectra enable model-independent deconvolution to obtain the parent electron spectrum. 
Taking into account the effects of albedo, these show that the low energy cut-off to the electron power-law spectrum is typically $\lapprox$tens of keV, confirming that the accelerated electrons contain a large fraction of the energy released in flares.\index{albedo}
\textit{RHESSI} has detected a high coronal hard X-ray source that is filled with accelerated electrons whose energy density is comparable to the magnetic-field energy density. 
This suggests  an efficient conversion of energy, previously stored in the magnetic field, into the bulk acceleration of electrons.  
A new, collisionless (Hall) magnetic reconnection process has been identified through theory and simulations, and directly observed in space and in the laboratory; it should occur in the solar corona as well, with a reconnection rate fast enough for the energy release in flares.
The reconnection process could result in the formation of multiple elongated magnetic islands,  that then collapse to bulk-accelerate the electrons, rapidly enough to produce the observed hard X-ray emissions. 
\textit{RHESSI's} pioneering $\gamma$-ray line imaging of energetic ions, revealing footpoints straddling a flare loop arcade, has provided strong evidence that ion acceleration is also related to magnetic reconnection.\index{arcade!and $\gamma$-ray images}
Flare particle acceleration is shown to have a close relationship to impulsive Solar Energetic Particle (SEP) events observed in the interplanetary medium, and also to both fast coronal mass ejections and  gradual SEP events. 
New instrumentation to provide the high sensitivity and wide dynamic range hard X-ray and $\gamma$-ray measurements, plus energetic neutral atom (ENA) imaging of  SEPs above $\sim$2~R$_\odot$, will enable the next great leap forward in understanding particle acceleration and energy release is large solar eruptions -- solar flares and associated fast coronal mass ejections (CMEs). 
\end{abstract}


\setcounter{tocdepth}{8} 
\tableofcontents

\section{Introduction}\label{sec:intro}

Large solar flares are the most powerful explosions in the solar system, releasing up to $10^{32-33}$~ergs in $10^{2-3}$~s. 
They and their associated fast coronal mass ejections (CMEs) are the most energetic particle accelerators in the solar system, producing ions up to tens of GeV and electrons to hundreds of MeV. For flares, the accelerated particles often appear to contain the bulk of the total energy released in the flare, a remarkable efficiency that indicates that the particle-acceleration and energy-release processes are intimately related. 
Much of the observed flare impulsive-phase phenomena appear to be the result of the interaction of these accelerated particles with the ambient medium. 
Fast CMEs drive shock waves that accelerate solar energetic particles (SEPs) observed near 1~AU with efficiency of order 10\%.
Similar processes are believed to occur elsewhere in the universe, in stellar flares 
\citep[e.g.,][]{2007ApJ...654.1052O}, magnetars \citep[e.g.,][]{2005Natur.434.1098H}, young circumstellar disks, supernovae shock waves, etc. 
Solar flares and CMEs are the most accessible laboratories for understanding the fundamental physics of transient energy release and efficient particle acceleration in cosmic magnetized plasmas.
Furthermore, these large solar eruptions produce the most extreme forms of space weather -- the radiation hazard from the most intense SEP fluxes, and the disruption of the the heliospheric plasma environment.\index{space weather!and extreme events}

The first observation of a solar flare was made by \cite{carrington1859} in white light.\index{white-light flares}
\index{flare (individual)!SOL1859-09-01T11:18 (pre-\textit{GOES})}
The first evidence that the Sun could accelerate particles to high energies came from the detection of a ground-level event (GLE)\index{cosmic rays!ground-level events} in a cosmic-ray sensor, reported by \cite{1946PhRv...70..771F}, who noted that it occurred near the time of a solar flare. 
Optical studies showed that flares typically occurred near sunspots, in regions of strong magnetic field, consistent with the release of energy stored in magnetic fields \citep[see][]{1948MNRAS.108..163G}.\index{sunspots!and flare occurrence}
Swept-frequency radio observations provided evidence for the acceleration and escape of fast electrons (Wild et al. 1963) in flares.\nocite{1963ARA&A...1..291W}
Soft X-ray (SXR) emission from a solar flare, indicating the presence of hot, $\sim$10$^7$~K thermal plasma, was discovered by \cite{1957Natur.179..861C}. 
Flare hard X-ray (HXR) emission was first reported by \cite{1958PhRvL...1..205P}, and $\gamma$-ray line emission by \cite{1973Natur.241..333C}; these radiations are generated through bremsstrahlung by energetic electrons and nuclear interactions of energetic ions, respectively, colliding with the ambient solar atmosphere. 
Since the cross-sections are known, and the solar atmosphere is optically thin to these energetic emissions, quantitative information about the parent electrons \citep[see][]{Chapter7} and ions \citep[see][]{Chapter4}, can be derived from the HXR and $\gamma$-ray line measurements.

For non-thermal particles with energy $E$~much greater than $kT$ (the average thermal energy of the ambient medium), the energy lost to Coulomb collisions is many orders of magnitude greater than the energy lost to HXRs and/or $\gamma$-rays. 
Assuming the particles lose all their energy to Coulomb collisions (collisional thick-target), the energy going into the source electrons and ions can be directly determined from the observed HXR and $\gamma$-ray line emissions \citep[see][]{Chapter3}. 
In the idealized situation where the acceleration of the particles can be considered separate (e.g., in the tenuous corona) from their loss to collisions (in the dense chromosphere and photosphere), the spectra of the accelerated electrons and ions can be inferred from the observed HXR and $\gamma$-ray spectra, respectively. 
Typically, the flux spectra are fit to power-laws in energy, $F(E) = AE^{-\delta}$, where $A$~and $\delta$ are constants;  the total energy in electrons and in ions depends critically on how low in energy the power-law extends.
This is the low-energy cutoff or roll-off parameter. 
For power laws extending down to $\sim$20~keV for the electrons, and down to $\sim$1~MeV for the ions, the accelerated electrons and ions each would contain a significant fraction, $\sim$10-50\%, of the total energy released in flares \citep{1976SoPh...50..153L}.   
The rate of acceleration of tens-of-keV electrons required in a large flare would be $\sim$10$^{36}$~s$^{-1}$, 
an enormous number equivalent to an impossibly large current of $\sim$10$^{17}$~A. 
This has led to models suggesting many filaments with oppositely directed currents, or with return currents \citep[see][]{Chapter8}, or electron acceleration directly in the dense chromosphere \citep[e.g.,][]{2008ApJ...675.1645F}.\index{filaments!current systems}

With the poor energy resolution of early flare HXR measurements, the spectra could be fit to a thermal spectrum with $T \approx 10^{8-9}$~K \citep{1978ApJ...223..620C}. 
Then $E \approx kT$, and Coulomb collisions would primarily exchange energy between electrons of comparable energy.
 In principle, there could be little or no net collisional loss, and the flare energy going into accelerated electrons could be orders of magnitude less than the total energy released in the flare. 
Flare HXR imaging has shown, however, that most of the HXR emission came from footpoints, co-spatial with chromospheric H$\alpha$ brightenings and other low-temperature flare emissions, so $E \gg kT$  \citep[see][]{Chapter2}. 
\index{footpoints!HXR and H$\alpha$}
Furthermore, the first measurements with high spectral resolution showed that the hottest thermal plasmas had $T \leq 40$~MK \citep{1981ApJ...251L.109L}, so the electrons producing HXR emission in coronal sources above $\sim$20 keV are mostly non-thermal
($E \gapprox 5 kT$).

\begin{figure}
\centering
\includegraphics[width=0.8\textwidth]{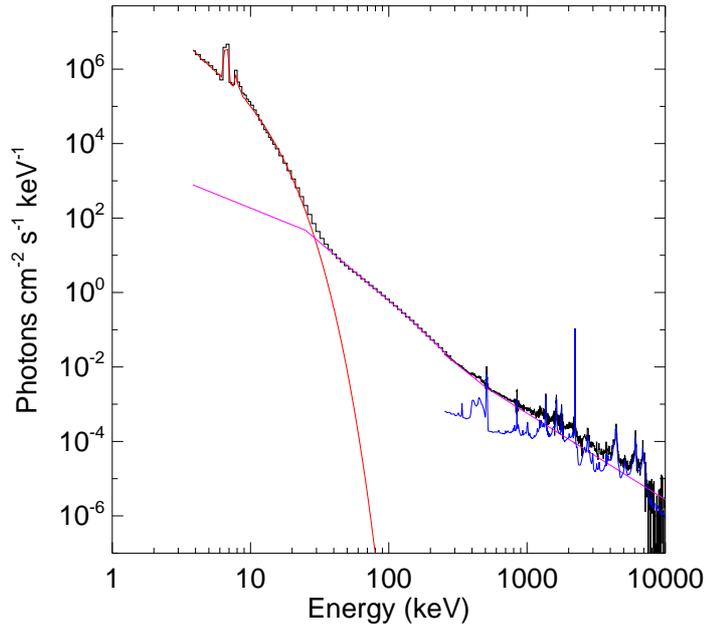}
\caption[]{\textit{RHESSI} measurement of the energy spectrum of SOL2002-07-23 (X4.8) from 3~keV to 10~MeV.
At energies below $\sim$30~keV, the emission is dominated by thermal plasmas with temperatures up to $\sim$40~MK, while accelerated electrons produce the spectrum detected above $\sim$30~keV. 
Narrow and broad $\gamma$-ray line emissions produced by accelerated ions are observed from $\sim$0.5 MeV to $\sim$8 MeV (blue line). 
In the hard X-ray range, the red line shows the thermal component, and the magenta line the HXR power-law component (with cutoff).
}
\label{fig:lin_fig1}
\index{flare (individual)!SOL2002-07-23T00:35 (X4.8)!illustration},
\end{figure}

\textit{RHESSI}, the \textit{Reuven Ramaty High Energy Solar Spectroscopic Imager} mission \citep{2002SoPh..210....3L} was designed to investigate how the Sun releases the energy for flares, presumably stored in the magnetic fields of the corona, and how electrons and ions are rapidly accelerated to high energies with such high efficiency. 
\index{satellites!RHESSI@\textit{RHESSI}}\index{RHESSI@\textit{RHESSI}}
\textit{RHESSI} provides high resolution imaging and spectroscopy of the X-rays and $\gamma$-rays that are emitted by the energetic electrons and ions in the flare. 
The measurements (Figure~\ref{fig:lin_fig1}) span almost four orders of magnitude in photon energy and more than 12~orders in flux, from the intense soft X-rays (SXRs) produced by hot thermal plasmas, through the HXRs emitted by accelerated electrons, to $\gamma$-ray line emission by accelerated ions\index{spectrum!gamma@$\gamma$-rays}. 
The cryogenically cooled germanium detectors (GeDs)\index{GeDs} provide uniquely high spectral resolution: $\sim$1~keV FWHM (full width at half maximum) up to $\geq 200$~keV, increasing to $\sim$2~keV at the 511~keV positron-annihilation line and $\sim$7~keV at 8~MeV, sufficient to resolve all the nuclear $\gamma$-ray lines except the 2.223~MeV neutron-capture deuterium line ($\leq 0.1$~keV FWHM intrinsic width). 
\index{FWHM}\index{neutrons!deuterium line}
\textit{RHESSI} provides HXR and $\gamma$-ray imaging (the first above 100~keV and the first ever for $\gamma$-ray lines) with the finest spatial resolution ever achieved ($\sim 2''$ from 3 to $\sim$100~keV, $\sim4''$ up to $\sim$300~keV, $\sim35''$ up to 17~MeV), utilizing rotating modulation collimators (RMCs) that convert the spatial information into a temporal modulation of the count rates.
\index{rotating modulation collimator}

\textit{RHESSI} observations span a broad range of solar flare (and quiet Sun) phenomena, described in the other chapters of this book. 
Here we summarize those results relevant to the fundamental processes of energy release and particle acceleration in flares, and synthesize them into a present understanding of the physics of flares. 
Some of the key questions addressed here are: 
How much energy is contained in the flare-accelerated electrons and ions?  
Where and how are the electrons and ions accelerated?  
Where and how is the energy released? 
What is the relationship of particle acceleration in flares related to the impulsive SEP events observed in the interplanetary medium?
\index{solar energetic particles (SEPs)}
How are large flares related to fast CMEs and the acceleration of SEPs in large gradual events?  
Finally, we then discuss future research directions and prospects. 
\index{key questions}

\section{Flare Acceleration of Electrons}

\subsection{The electron spectrum}

As discussed in \cite{Chapter7}, the observed flare HXR emission depends on the source electron spectrum, convolved with the bremsstrahlung cross-section, multiplied by the ambient density, and integrated over the line of sight.  
In the past, typically a source electron spectral shape was assumed, and then forward-fit to the HXR observations. 
\cite{1992SoPh..137..121J} showed that the HXR continuum measurements can be directly deconvolved in a model-independent way to obtain the source electron spectrum (but crucially dependent on the statistics of the observations); they obtained source electron spectra for the HXR measurements of SOL1980-06-07T03:22 (M7.3)\index{flare (individual)!SOL1980-06-07T03:22 (M7.3)!modeling}, the first hard X-ray flare observed with high spectral resolution \citep{1993ApJ...417L..53L}. 

For intense solar flares, \textit{RHESSI} provides excellent statistics together with $\sim$1~keV FWHM resolution for the most precise HXR continuum measurements ever obtained from an astrophysical source. 
This has enabled routine, model-independent deconvolution of the HXR spectrum by powerful newly-developed mathematical techniques 
\citep[e.g., regularized inversion; see][]{Chapter7} to obtain the spectrum of the source electrons. 
The observed HXR spectrum also depends on the angular distribution of the source electrons, and on the Compton scattering of the source HXRs by the solar photosphere, i.e., albedo.\index{albedo}
The latter can be significant, especially at deka-keV energies.
An elegant Green's function approach to evaluating the albedo contribution that is independent of the primary spectrum has been developed \citep{2006A&A...446.1157K}.\index{albedo}\index{Green's functions}

The number of electrons and the total energy they contain depends critically on the low-energy cutoff.
\index{accelerated particles!low-energy cutoff}\index{electrons!spectrum!low-energy cutoff}
\textit{RHESSI's} spectral resolution is sufficient to resolve the steep high energy fall-off of hot flare thermal continuum (e-folding of $\sim$2~keV), allowing the precise determination of the energy above which the HXR emission must be non-thermal \citep[see][]{Chapter3}.  
The newly-developed deconvolution methods have been applied \citep{2005SoPh..232...63K} to flares with non-thermal HXR spectra that showed flattening at low energies (as expected for a cutoff or roll-off in the electron source spectrum spectra). 
For this initial sample of \textit{RHESSI}-observed flares, the derived electron source spectra appeared to have a roll-off around 20-40~keV, but all these flares were located close to the solar disk center where the effects of albedo are strongest. 
After correcting for the expected albedo, all the derived source electron spectra extend in a power law with no roll-off down to where the hot flare thermal emission dominates, typically $\leq$20~keV and sometimes as low as $\sim$12~keV for these flares \citep{2008SoPh..252..139K}.
\index{low-energy cutoff} 
Low-energy cutoffs were also found for a number of other flares \citep{2005ApJ...626.1102S,2007ApJ...670..862S} in the $\sim$15-50~keV range, implying that the non-thermal electrons indeed must contain a large fraction of the energy released in many flares. 

\subsection{Early phase}

\begin{figure}
\centering
\includegraphics[width=0.8\textwidth]{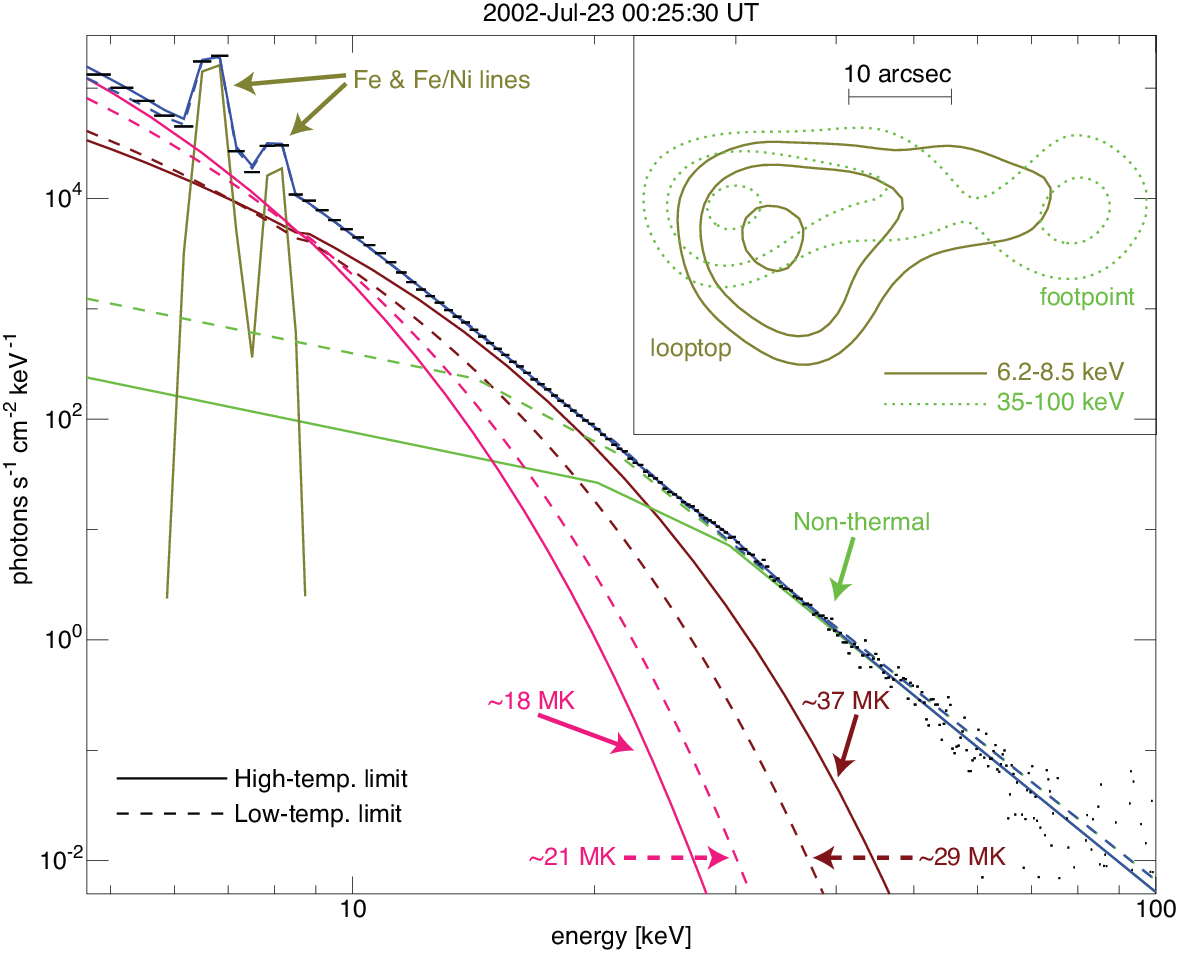}
\caption[]{Photon flux spectra (black) during the peak of the pre-impulsive phase of SOL2002-07-23 ($\sim$00:25:30~UT), with two acceptable model fits showing the upper (solid) and lower (dashed) temperature limits of the super-hot component (brick), as constrained by the Fe~and Fe/Ni lines (olive). 
Cool isothermal (magenta) and non-thermal (green) components are also required. 
(Inset) 30\%, 50\%, and 90\% contours of 6.2-8.5~keV (black solid) and 35-100~keV (red dotted) images at the same time. 
The peak non-thermal emission appears to be above the thermal looptop; the faint footpoint contains only $\sim$16\% of the total non-thermal flux within the 50\% contour \protect\citep[from][]{2010ApJ...725L.161C}.
}
\label{fig:lin_fig6}
\index{flare (individual)!SOL2002-07-23T00:35 (X4.8)!illustration}
\index{flare (individual)!SOL2002-07-23T00:35 (X4.8)!superhot component}
\index{flare (individual)!SOL2002-07-23T00:35 (X4.8)!pre-impulsive phase}
\end{figure}

In the X4.8 flare SOL2002-07-23T00:35\index{flare (individual)!SOL2002-07-23T00:35 (X4.8)!hard X-ray precursor} (and several other large flares) \textit{RHESSI} detected a weak HXR source high in the corona beginning $\sim$9~min before the impulsive phase, and prior to the detection of any footpoint HXR emission \citep{2002SoPh..210....3L}, or of any chromospheric counterpart in \textit{TRACE} 195~\AA, \textit{SOHO}/MDI white light, or H$\alpha$ \citep{2003ApJ...595L.103K}\index{pre-impulsive phase}.\index{precursor!hard X-ray}
The lack of footpoint emission indicates that the thermal source has its origin in the corona and does not come from chromospheric evaporation. 
Such a weak source could not have been detected by previous solar HXR instruments that had fixed windows optimized for the peak emission of large flares; \textit{RHESSI} inserts attenuators to absorb low energy X-rays as the count rate increases, so its sensitivity is much higher when one or both of the attenuators are out. 
Both thermal (superhot)\index{superhot component} plasma and non-thermal electrons are detected (Figure~\ref{fig:lin_fig6}), with a much steeper (softer) energy spectrum than for electrons accelerated in the flare impulsive phase. 
As the pre-impulsive phase progresses, the coronal source appears to move downward, as expected when stored magnetic energy is being released by shortening of magnetic field lines.  
Some footpoint HXR emission appears later, but generally at much lower intensity than the coronal HXR source. 
In this phase only a small fraction of total flare energy is released. 
This pre-impulsive source may be due in part to the initial energization by magnetic reconnection process itself (to be discussed later).  

\subsection{Energy release}

\begin{figure}
\centering
\includegraphics[width=\textwidth]{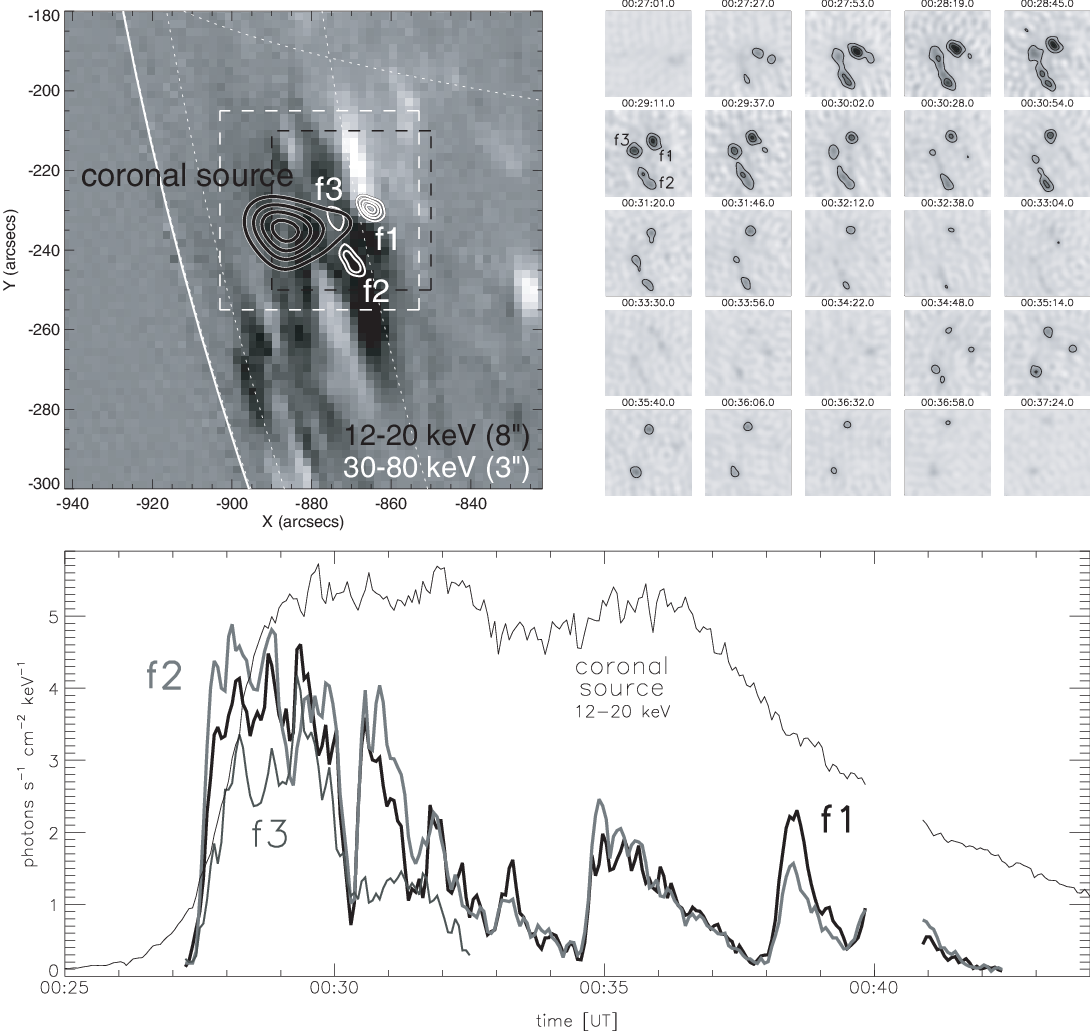}
\caption[]{Location and temporal variation of the different HXR sources observed in the $\gamma$-ray flare SOL2002-07-23 (X4.8). 
\textit{Top left:} the three main 30-80 keV HXR footpoint sources (white contours) and the thermal plasma (12-20 keV, black contours), superimposed on the pre-flare magnetogram (MDI). 
\textit{Top right:} 30-80 keV images over the area indicated by the black dashed line in the top left panel, taken with 26~s integration time, scaled to the maximum intensity of the time series.
\textit{Bottom:} flux-vs-time profiles of the three HXR (30-80 keV) footpoints, plus the coronal thermal source (thin curve, 12-20 keV flux divided by 1500) from images taken every 4~s \protect\citep[from][]{2003ApJ...595L.103K}.
}
\label{fig:lin_fig2}
\end{figure}

\begin{figure}
\centering
\includegraphics[width=0.43\textwidth]{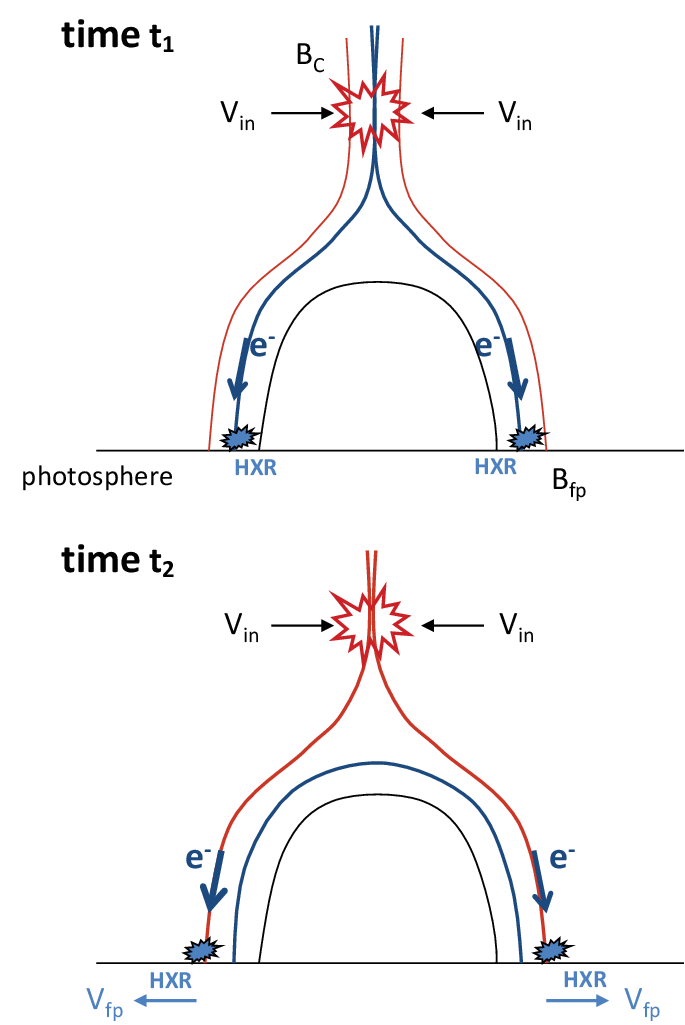}
\includegraphics[width=0.56\textwidth]{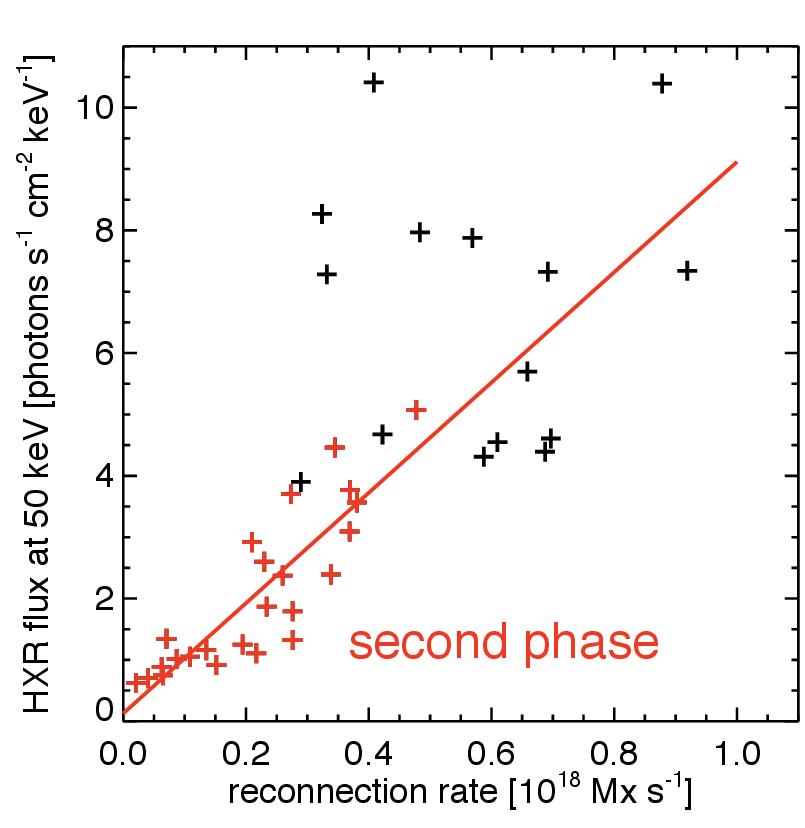}
\caption[]{\textit{Left:} schematic of reconnecting field lines in the corona, showing the 
	relationship to measurements in the chromosphere/photosphere.
	\textit{Right:} the observed hard X-ray flux at 50~keV plotted versus the rate of magnetic flux being reconnected for SOL2003-10-29 \protect\citep[from][]{2005AdSpR..35.1707K}.
}
\label{fig:lin_fig3}
\index{flare (individual)!SOL2003-10-29T20:49 (X10.0)!illustration}
\end{figure}

During the impulsive phase, the rate $d\Phi/dt$ of magnetic flux $\Phi$~being reconnected in the corona can be inferred from the observed apparent velocity, $v_{fp}$, of the HXR footpoints (assumed to be magnetically connected to the photosphere just below; see Figure \ref{fig:lin_fig2}) and the measured photospheric magnetic field, $B_{fp}$.
The flux change $d\Phi/dt = v_{c}B_ca_c$ must be equal to the $v_{fp}B_{fp}a_{fp}$, if the magnetic field is frozen to the plasma (see Figure~\ref{fig:lin_fig3}, left).\index{frozen-in field}
Here $a_{fp}$ is the footpoint width perpendicular to its motion, as given by the imaging, and $v_c$, $B_c$, and $a_c$ are the velocity, magnetic field, and width in the corona respectively. 
Note that in three dimensions the convection electric field in the corona is $E_c = v_cB_c = v_{fp}B_{fp} (a_{fp}/a_c)$, as opposed to $E_c = v_cB_c = v_{fp}B_{fp}$  for the 2-D case.\index{electric fields!convective} 
Figure~\ref{fig:lin_fig3} (right) shows that the HXR flux at 50 keV is roughly proportional to the measured reconnection rate $v_{fp}B_{fp}a_{fp}$ for the X10 flare SOL2003-10-29T20:49 (X10.0)\index{flare (individual)!SOL2003-10-29T20:49 (X10.0)!acceleration} \citep{2005AdSpR..35.1707K}.  
The HXR flux is also roughly correlated with $v_{fp}B_{fp}^2$ \citep[see Figure~3.12 of][]{Chapter2}, consistent with models where a significant fraction of the magnetic energy released by reconnection goes to accelerating electrons. 
The correlation is better after $\sim$20:48~UT (red symbols in Figure~\ref{fig:lin_fig3}, right) when the geometry is close to the simplest 2-D reconnection model with just two HXR sources; earlier in the flare, the HXR emission is complex with at least two sources on each flare ribbon\index{hard X-rays!ribbons}\index{ribbons!hard X-ray}, and the scatter of points is much greater. 
For this flare the convection electric field $E_c = v_cB_c$ is as large as 6000($a_{fp}/a_c$)~V~m$^{-1}$, and the reconnection rate reaches as high as $d\Phi/dt \sim 10^{18}$~Mx~s$^{-1}$.
\index{electric fields!convective}
Reconnection rates of the same order of magnitude are obtained for other large flares, together with similar rough linear correlation with the flare HXR fluxes. 
$E_c$  is generally perpendicular to the magnetic field, and it should map down to the chromosphere, but if the reconnecting fields are not anti-parallel, there will be a component of $E_c$ parallel to $B$ in the reconnection region. 
These strong electric fields may  be important for particle acceleration.

\subsection{Energy deposition}

Where does the energy deposited by the non-thermal electrons into the solar atmosphere go? As discussed in \cite{Chapter2}, energy deposition by energetic electrons at the top of the chromosphere produces  hot thermal flare plasma through the process of chromospheric evaporation. 
Heating lower in the atmosphere is expected to result in very substantial amounts of radiation.  
The \textit{SORCE}\index{satellites!SORCE@\textit{SORCE}} mission provided the first measurements precise enough to detect increases in total solar irradiance (TSI)\index{TSI} for flares.  For the very large X17 flare SOL2003-10-28\index{flare (individual)!SOL2003-10-28T11:10 (X17.2)!TSI increase}, \cite{2004GeoRL..3110802W} found an increase in TSI  with a time profile that shows an impulsive component similar to the HXRs, as well as a gradual component similar to the soft X-rays.  
The maximum of the TSI increase ($\sim$10$^{30}$~erg~s$^{-1}$) is of the same order of magnitude as that expected from the energy deposited by the non-thermal electrons, and it occurs at the impulsive-component peak, prior to the SXR peak. 
Thus, at the present time, the observations are consistent with a large fraction of the total energy in a flare being released in the form of accelerated electrons (and ions) during the impulsive phase; these then heat the ambient atmosphere through collisions, evaporating chromospheric material  to form the hot (above tens of MK) flare plasma, and heating the deeper solar atmosphere with the energy escaping as radiation.  
Most other impulsive phase flare phenomena (such as radio emission, etc.) can be explained as the consequence of the interaction of the accelerated particles with the ambient medium.

The HXR spectra typically fit a double powerlaw with a downward break at an energy of tens of keV, reaching up to $\geq$100~keV in large flares. 
This break is often surprisingly sharp, but not unphysically so \citep{2003A&A...407..725C}. 
As discussed in \cite{Chapter3}, the break could be due to non-uniform ionization of the atmosphere (from the fully ionized corona to the neutral chromosphere) where the electrons lose their energy to collisions, or due to return-current energy loss or other wave-particle interactions, or due to the acceleration process itself.\index{non-uniform ionization} 
In impulsive solar electron events observed at $\sim$1~AU, the electron energy spectra are also observed to be a double powerlaw with a break \citep{2007ApJ...663L.109K}.

\textit{RHESSI} also provides Fourier-transform HXR imaging with the finest spatial resolution ever achieved ($\sim$2$''$ from 3 to $\gapprox$100~keV).  
Figure~\ref{fig:lin_fig2} (top right) shows a temporal sequence of HXR images for SOL2002-07-23T00:35 (X4.8)\index{flare (individual)!SOL2002-07-23T00:35 (X4.8)!footpoints}, revealing three footpoints that are cospatial with flare H$\alpha$ brightenings \citep{2004ApJ...611..557A}.
This implies non-thermal electrons as an energy source.
The bottom panel shows that the HXR flux variations with time are generally the same for each of the three footpoints, with no obvious delays (within the few-second temporal resolution) of the peaks from one footpoint to another. 
\cite{2000AdSpR..26..497S} cross-correlated the HXR fluxes in paired footpoints of other flares observed by \textit{Yohkoh}, and showed that the HXR emissions were simultaneous to within a few tenths of a second.\index{Yohkoh@\textit{Yohkoh}!HXT}
These observations are consistent with the ``standard'' CSHKP\footnote{Carmichael, Sturrock, Hirayama, Kopp \& Pneuman.}
\index{flare models!CSHKP}\index{flare models!standard}\index{satellites!Yohkoh@\textit{Yohkoh}}\index{CSHKP}
model of solar flares \citep[e.g.,][]{Chapter2}, where magnetic reconnection in roughly oppositely-directed magnetic fields above the closed loops forms a new loop (Figure~\ref{fig:lin_fig3}, left) in the corona and accelerates electrons which, because of their high speed ($\gapprox$$0.3 c$), stream down the new loop and arrive at the two footpoints essentially simultaneously.   
This is a stringent constraint on models where the electron acceleration occurs in the chromosphere, as advocated by \cite{2008ApJ...675.1645F}; their Alfv{\' e}n waves (proposed to transport the energy to the chromosphere) must have  velocities comparable to  $\sim$$0.3 c$.\index{waves!Alfv{\' e}n!and energy transport}\index{transport!in Alfv{\' e}n waves}\index{acceleration!chromospheric}

\subsection{Electron acceleration region}
\index{acceleration region!electrons}

\textit{RHESSI's} HXR imaging (and previous measurements) are generally consistent with the collisional thick-target model\index{flare models!thick-target} \citep[see Section~7,][]{Chapter3}.
Many lines of reasoning point to the corona as the location of the flare energy release and particle acceleration sites, and the coronal magnetic field as the source of the energy that powers flares. \textit{RHESSI} sometimes detects two relatively weak coronal X-ray sources with opposing energy gradients \citep[see Figure~4.2,][]{Chapter2}, one above the other \citep{2003ApJ...596L.251S,2008ApJ...676..704L}, implying the energy release site lies between them at altitudes of $\sim$9-23~Mm, as expected for a current sheet formed between the top of flare loops and the coronal source.\index{current sheets}
Using high-sensitivity HXR observations from the BATSE instrument on the \textit{Compton Gamma-ray Observatory (CGO)}, \cite{1995ApJ...455..347A} found a delay of low-energy HXRs for short-duration bursts, the delays being consistent with time-of-flight for the parent electrons from a high coronal source down to the chromospheric footpoints.  
Furthermore, several analyses of the HXR footpoint height as a function of energy show that the centroids of the emissions are at lower altitudes for higher energies, consistent with the parent electrons being injected from above the chromosphere \citep[e.g.,][]{2002SoPh..210..383A}. 
In the well-known \cite{1994Natur.371..495M} flare, a HXR source was detected above the soft X-ray looptop. 
With the limited dynamic range of \textit{Yohkoh}/HXT and \textit{RHESSI}, however, weak coronal HXR sources are difficult to detect in the presence of the much brighter footpoint sources, but systematic studies of flares whose footpoints are occulted show that coronal HXR emission is commonly present \citep{2008ApJ...673.1181K}. 
\index{satellites!Yohkoh@\textit{Yohkoh}}\index{Masuda flare}
\index{hard X-rays!dynamic range}\index{occulted sources}

\begin{figure}
\centering
\includegraphics[width=\textwidth]{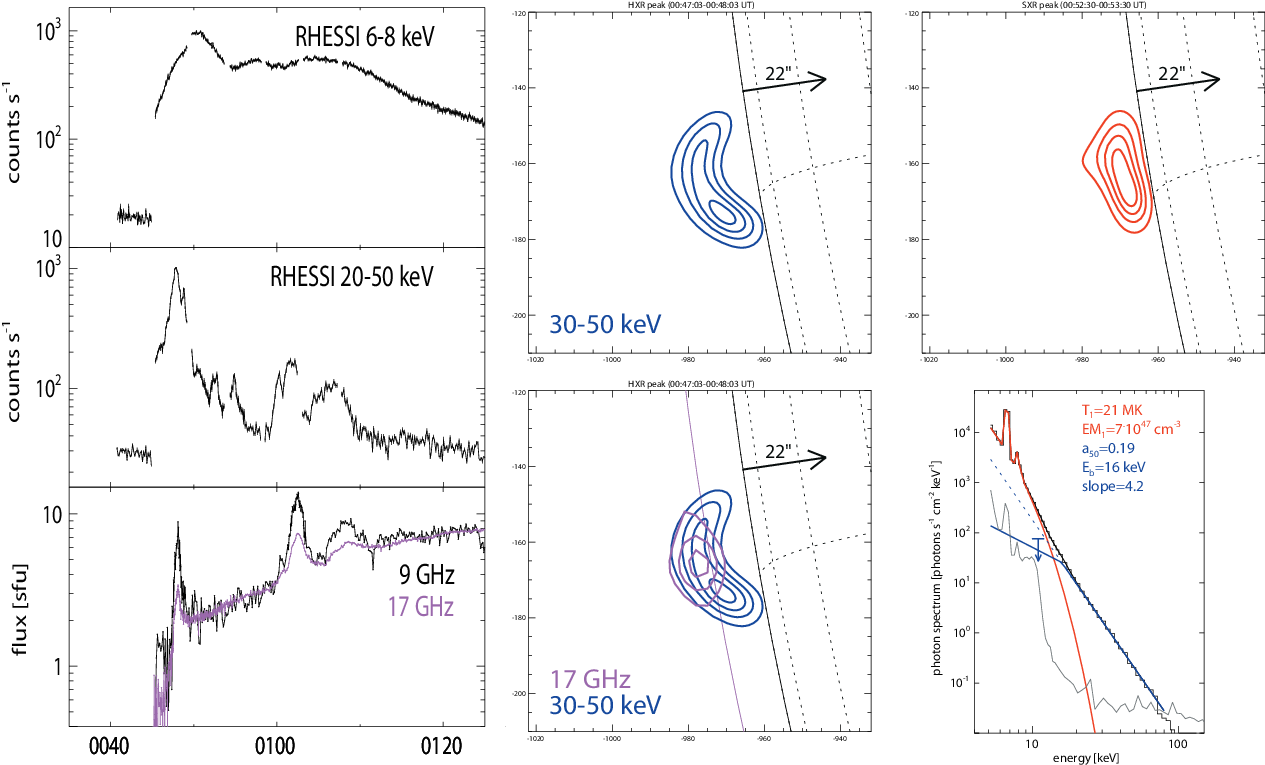}
\caption[]{HXR (\textit{RHESSI}) and radio (Nobeyama) observations of coronal emission from the over-the-limb flare SOL2007-12-31T01:11 (C8.3).   
\textit{Left}: flux vs. time profiles of thermal 6-8~keV X-rays (top panel), 20-50 keV HXRs (middle panel),  9 and 17~GHz radio (bottom panel). 
\textit{Middle}: top, a 30-50~keV HXR image with limb indicated; bottom, a 17~GHz radio image (purple) with the radio limb indicated, superimposed on the HXR image.
\textit{Right}: top, Image of the soft X-ray source located below the HXR source; bottom, burst HXR energy spectrum (blue) with pre-burst SXR thermal spectrum (red).
}
\label{fig:lin_fig4}
\end{figure}

More recently, \textit{RHESSI} has detected an intense coronal HXR source located about 6~Mm above the soft X-ray flare loops (Figure~\ref{fig:lin_fig4}, middle and right upper panels) in SOL2007-12-31T01:11 (C8.3), where the footpoints were occulted, but with \textit{STEREO~B} providing un-occulted EUV imaging of the whole flare \citep{2010ApJ...714.1108K}. 
\index{coronal sources}\index{occulted sources}
\index{flare (individual)!SOL2007-12-31T01:11 (C8.3)!coronal hard X-ray source} 
This coronal HXR source shows several impulsive bursts at 20-50~keV (Figure~\ref{fig:lin_fig4}, left middle panel), with a power-law spectrum extending to $>$80~keV, with no detectable thermal emission (Figure~\ref{fig:lin_fig4}, lower right panel). 
Thus, essentially all the electrons in this source have been accelerated.  
The HXR imaging shows a source of volume $\sim 10^{27}$~cm$^3$, with a non-thermal, $>$16~keV electron density of $\sim 2 \times 10^9$~cm$^{-3}$ for a total of $N(> 16\ \mathrm{ keV}) \approx 2 \times 10^{36}$ electrons in the source (see Table~\ref{tab:lin_tab1}, center column). 

\index{flare (individual)!SOL2007-12-31T01:11 (C8.3)!illustration}

\begin{table}
\begin{center}
\index{coronal sources!table of parameters}
\index{flare (individual)!SOL2007-12-31T01:11 (C8.3)!table of parameters}
\index{flare (individual)!SOL2003-10-22T20:07 (M9.9)!table of parameters}
\index{flare (individual)!SOL1992-01-13T17:25 (M2.0)!table of parameters}
\caption{Coronal Hard X-ray Sources}
\label{tab:lin_tab1}
\smallskip
\begin{tabular}{| l | c | c | c |}
\hline
Parameter & SOL1992-01-13 & SOL2007-12-31 & SOL2003-10-22 \\ \hline
                    & (Masuda flare)$^a$ & & (peak 3) \\ \hline
\hline
Burst duration & $\sim$2 min & $\sim$2 min & $\sim$2 min \\ \hline
Fastest decay (1/e) time &  $\sim$30 s&  $\sim$30 s&  $\sim$30 s \\ \hline
Energy range &$\sim$25-50 keV&$\sim$16-80 keV&$\sim$30-80 keV\\ \hline
Flux at 50 keV & $\sim$0.02& $\sim$0.2& $\sim$0.2\\ 
ph~(cm$^2$ s keV)$^{-1}$ &&&\\ \hline
Height above photosphere & $\sim$20 Mm & $\sim$27 Mm & $\sim$25 Mm \\ \hline
Height above flare loop & $\sim$7 Mm & $\sim$6 Mm & $\sim$6 Mm \\ \hline
Length& $\sim$5 Mm & $\sim$29 Mm & $\sim$11 Mm \\ \hline
Width& $\sim$5 Mm & $\sim$6 Mm & $\sim$6 Mm \\ \hline
Volume$^b$& $\sim1 \times 10^{26}$~cm$^{3}$ &$\sim8 \times 10^{26}$~cm$^{3}$ & $\sim4 \times 10^{26}$~cm$^{3}$ \\ \hline
Pre-flare ambient density$^c$& low & low & low \\ \hline
Electron spectral index & $\sim$3-4.5 &  $\sim$3.7 &  $\sim$4.6 \\ 
(thin-target hard X-rays) &&& \\ \hline
Non-thermal electron density & $\sim2 \times 10^{9}$~cm$^{-3}$ &$\sim2 \times 10^{9}$~cm$^{-3}$ & $\sim1 \times 10^{10}$~cm$^{-3}$ \\ \hline
Number of electrons $>$16~keV& $\sim2 \times 10^{35}$ &$\sim2 \times 10^{36}$ & $\sim4 \times 10^{36}$ \\
\hline
Energy, non-thermal electrons$^d$ & $>1 \times 10^{28}$ erg & $>1 \times 10^{29}$ erg & $>3 \times 10^{29}$ erg \\ \hline
Radio flux at 17~GHz & $\sim$600 SFU& $\sim$1.7 SFU & \\ \hline
Electron spectral index (radio) & & $\sim$3.4 & \\ \hline
Typical electron energy & & $\sim$1.2~MeV & \\
(17 GHz emission)  & & & \\ \hline
Magnetic field strength & & $\sim$30-50 G & \\ \hline
Pre-flare plasma $\beta$$^e$ & & $\sim$0.01 & \\ \hline
Plasma $\beta$ in HXR source & & $\sim$1 & \\ \hline
Footpoint X-ray flux at 50 keV & $\sim$0.1 ph~(cm$^2$ s keV)$^{-1}$ & & $\sim$0.1  ph~(cm$^2$ s keV)$^{-1}$\\ \hline
Footpoint area & & & 12 Mm$^2$\\ \hline
Footpoint electron spectral & $\sim$4.2-5.0 & & $\sim$4.6 \\ 
index (thick target) &&& \\ \hline
Footpoint electron loss rate & $\sim4 \times 10^{35}$~e$^-$~s$^{-1}$& & $\sim1 \times 10^{35}$~e$^-$~s$^{-1}$\\ 
($>$16 keV) &&& \\ \hline
Footpoint energy deposition & $>2 \times 10^{28}$  erg~s$^{-1}$ & & $>6 \times 10^{27}$  erg~s$^{-1}$\\ 
rate ($>$16 keV) &&& \\ \hline
Footpoint energy flux &&& $>5 \times 10^{10}$ erg~(cm$^2$ s)$^{-1}$ \\ \hline
\end{tabular}
\end{center}

$^a$Parameters taken from \cite{2000AdSpR..26..493M}\\
$^b$Volume = length $\times$ width $\times$ width\\
$^c$For the case that all electrons are accelerated, the pre-flare density is as given below\\
$^d$Derived assuming that all electrons are accelerated\\
$^e$Assumes a pre-flare temperature of 2~MK\\

\end{table}\index{Masuda flare!table}

Microwave imaging at multiple wavelengths (from Nobeyama) shows a co-spatial source (Figure~\ref{fig:lin_fig4}, lower middle panel), consistent with gyrosynchrotron emission from the high-energy tail (electron energies of $\sim$1~MeV) of the same power-law electron distribution.  
\index{flare (individual)!SOL2007-12-31T01:11 (C8.3)!microwave imaging}
The magnetic field in the source, estimated from the turnover in the radio spectrum, is $\sim$30-50~G, implying that the energy density of the non-thermal electrons in the source is comparable to that of the magnetic field, i.e., the plasma $\beta \approx 1$. 
Prior to the flare, no HXR emission was detected above background from this location; if we assume the density is the same as during the flare, the pre-flare plasma $\beta \approx 0.01$. 

\begin{figure}
\centering
\includegraphics[width=0.6\textwidth]{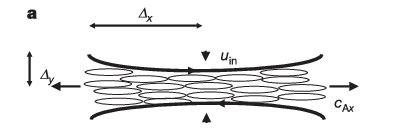}
\includegraphics[width=0.6\textwidth]{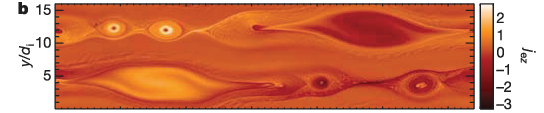}
\caption[]{(a) Diagram showing volume-filling islands expected around the field reversal region. 
(b) Particle-in-cell simulation of island formation during magnetic reconnection; shown here is the electron out-of-plane current at time t = 20~$\Omega_{ci}^{-1}$ where $\Omega_{ci}$ is the ion cyclotron frequency\index{frequency!Larmor!ion} \protect\citep[from][]{2006Natur.443..553D}.
\index{simulations!PIC!illustration}
}
\label{fig:lin_fig5}
\end{figure}

The impact of these observations on theoretical models of particle acceleration has been summarized in \cite{Chapter8}.
Particle-in-cell simulations of reconnection \citep{2006GeoRL..3313105D} show that narrow current layers form at the X-line and produce secondary magnetic islands (Figure~\ref{fig:lin_fig5}b).\index{simulations!PIC!acceleration in magnetic islands}\index{current sheets!and magnetic islands}
\cite{2006Natur.443..553D} argue that in three dimensions the islands should be volume-filling, and that electrons should be efficiently accelerated, primarily by the contraction of the initially squashed magnetic islands (Figure~\ref{fig:lin_fig5}b), e.g, converting elongated magnetic fields to more potential fields.
Some acceleration may also come from parallel electric fields in the reconnection process.\index{electric fields!parallel} 
\index{magnetic structures!islands!Fermi acceleration} 
Electrons circulating rapidly within the islands gain energy through a Fermi process\index{acceleration!Fermi!in shrinking islands}, by reflecting off the ends of the islands as they move inward at the Alfv{\' e}n speed; they interact with multiple islands to reach high energies and produce a powerlaw spectrum.\index{Alfv{\' e}n speed}
This process operates for pre-event $\beta \ll 1$.\index{Alfv{\' e}n speed!and reconnection}
The island contraction ceases when the energetic electron pressure approaches the local magnetic energy density, i.e., $\beta \sim 1$. 
Up to 60\% of the released magnetic energy can be transferred to the electrons in the process.  

As pointed out previously, a major issue is the supply of large numbers of electrons per second. 
The rate of loss of electrons from this source can be estimated by N($>$16~keV)/$\tau_{\mathrm{loss}}$, and ranges from a maximum of $5 \times 10^{37}$~electrons~s$^{-1}$ for $\tau_{\rm loss} \approx 0.4$~s (the time for a 16-keV electron to traverse the source) to a minimum of $\sim 3 \times 10^{35}$ electrons~s$^{-1}$ for $\tau_{\mathrm{loss}}  \approx 20$~s (the fastest observed HXR decay time). 
Assuming the reconnection inflow speed is $\sim 0.1 v_{\rm A}$ (discussed later), where $v_{\rm A}$  is the Alfv{\' e}n speed, then the rate that electrons are brought into the reconnection region from both sides can be estimated as $dN/dt = 2 \times 0.1 n_e A v_{\rm A}$, where $A$ is the cross-sectional area of the inflow stream, assumed to arrive from both sides.\index{Alfv{\' e}n speed!and reconnection}
For the parameters of this flare (Table~\ref{tab:lin_tab1}), $v_{\rm A} = 2.2 \times 10^8$~cm~s$^{-1}$, and 
$dN/dt \approx 8 \times 10^{35}$~electrons~s$^{-1}$.  
These numbers of electrons being convected into the acceleration region are sufficient to produce the footpoint HXR emission typical for this class (\textit{GOES} M2) flare, provided the bulk of the electrons are accelerated, as suggested by these observations. 
Thus, the number of electrons and the rate that they can be accelerated may be sufficient, in principle, to produce the footpoint HXR emission in a large flare; this above-the-looptop source can be identified as the acceleration region for the flare. 
\index{above-the-looptop sources!identification with acceleration region}

Another above-the-looptop coronal source was observed in HXRs together with one footpoint source for one of the bursts in the event SOL2003-10-22T20:07 (M9.9)! \citep{ishikawa_letter}.
\index{flare (individual)!SOL2003-10-22T20:07 (M9.9)!above-the-loop-top source} 
The energy spectrum of the electrons in the coronal source (inferred from the thin-target HXRs) has a power-law spectral index consistent with that required to produce the footpoint HXR (thick-target) emission as they lose their energy to collisions (see Table~\ref{tab:lin_tab1}, right column).\index{thin target!and coronal sources}
The rate (electrons~s$^{-1}$) required to produce the footpoint HXR emission would empty the coronal source in about 4~s, compared to the 2~min duration of the burst. 
For comparison, in the Masuda flare\index{flare (individual)!SOL1992-01-13T17:25 (M2.0)!time scales}
 the coronal source would be emptied in $\sim$0.5~s compared to the $\sim$2~min burst duration (Table~\ref{tab:lin_tab1}, left column). 
These examples are consistent with the bulk acceleration of electrons in the corona through a mechanism involving magnetic reconnection, with an inflow speed of $\sim$0.1~$v_{\rm A}$, but higher sensitivity and dynamic range are required to test whether these flares are unusual or whether this is common to all flares. 
The observed wide range of flares, however, suggest that perhaps a variety of possible mechanisms may be operating, ranging from energy release/particle acceleration in the coronal magnetic reconnection region, to particle acceleration in the chromosphere driven by Alfv{\' e}n waves from reconnection region \citep[as for the Earth's aurora;][]{2008ApJ...675.1645F}.\index{aurora}\index{flare models!auroral}

\section{Flare-accelerated Ions}

Information on the energy spectrum of the energetic ions is derived from ratios of $\gamma$-ray lines whose cross-sections have different energy thresholds.  
The handful of $\gamma$-ray lines that have been detected with sufficient statistics in flares provide information only for protons above $\sim$2.5~MeV. 
The inferred energy spectra of the accelerated ions, and the total energy contained in the ions, also depend on the composition and angular distributions of those ions; and on the composition, magnetic field, density, temperature, and scattering properties of the ambient medium -- thus multi-parameter models must be utilized and the results can vary by up to an order of magnitude \citep[see Section~2.1 of][]{Chapter4}. 
Assuming that the ion power-law spectra extend down to $\sim$1~MeV and the composition of the accelerated ions is the same as for impulsive SEPs\index{solar energetic particles (SEPs)!and flare energy} events observed at 1~AU (i.e., with strong enhancements of heavy ions), the total energy contained in flare-accelerated ions is found to be comparable to the energy in flare-accelerated electrons above $\sim$20 keV and thus a substantial fraction of the total energy released in the flare \citep{1995ApJ...455L.193R,2000AIPC..522..401R,2003ApJ...595L..69L}. 

\textit{RHESSI} has provided a major breakthrough with the first imaging of energetic ions in flares 
\citep[see Section~5 of][for details]{Chapter4}, using the strong, narrow 2.223~MeV neutron-capture line that is produced primarily by $>$30 MeV protons.\index{neutrons}
Five flares have been imaged to date \citep{2003ApJ...595L..77H,2006ApJ...644L..93H} and the $\gamma$-ray line emission is found to come entirely from compact unresolved ($\leq 35''$, \textit{RHESSI's} $\gamma$-ray angular resolution) sources located in the flare region, with upper limits of order $\sim$10\% for larger-scale diffuse emission. 
This clearly shows that the acceleration of the parent ions is predominantly flare-related and, for the most part, not due to acceleration by widespread shock waves \citep[but see][]{1993ApJ...409L..69V}, such as the fast CME-driven shocks that appear to accelerate the SEPs\index{shocks!particle acceleration}\index{solar energetic particles (SEPs)!shock acceleration} in large gradual SEP events observed near 1~AU.\index{ions!acceleration!in flares}\index{acceleration!by flares}

For SOL2003-10-28T11:10 (X17)\index{flare (individual)!SOL2003-10-28T11:10 (X17.2)!$\gamma$-ray imaging}, two $\gamma$-ray line sources are detected straddling the arcade of flare loops, strongly suggesting that the acceleration of ions is also associated with magnetic reconnection, similar to the acceleration of electrons.\index{arcade}
The two $\gamma$-ray line sources have about the same separation between footpoints and about the same orientation as the two $>$0.2~MeV HXR sources observed in the flare (contrary to expectations of stochastic acceleration models where ions are accelerated in larger loops; Emslie et al., 2004),\nocite{2004ApJ...602L..69E} but both $\gamma$-ray line sources are significantly displaced, by $\sim$15$''$, from the HXR sources. 
Gradient and curvature drifts in a simple loop arcade field would produce a displacement much smaller than observed, although in the same direction.\index{arcade!particle drifts}  
In the other four flares, two $\gamma$-ray line sources straddling the flare loops would not be resolvable, given the count statistics and flare loops widths. 
For SOL2002-07-23T00:35 (X4.8)\index{flare (individual)!SOL2002-07-23T00:35 (X4.8)!$\gamma$-ray imaging}
 the centroid of the $\gamma$-ray line source is displaced by $\sim$25$''$ from the HXR source centroid. 
For two other flares, the $\gamma$-ray line sources appeared to be associated with one of the two HXR footpoints, although statistics for one flare were marginal; for the fifth flare the statistics were too poor to tell. 
These differences in spatial morphology between electrons and ions are surprising, given the otherwise close correlation between the 2.223~MeV line fluences and the $>$0.3~MeV electron bremsstrahlung fluences (discussed below).
This may provide a clue to the acceleration and/or propagation of the two species, suggesting that electric fields may play a role \citep{1993SoPh..146..127L,2004ApJ...604..884Z}.\index{electric fields!and particle segregation} 

\begin{figure}
\centering
\includegraphics[width=0.54\textwidth]{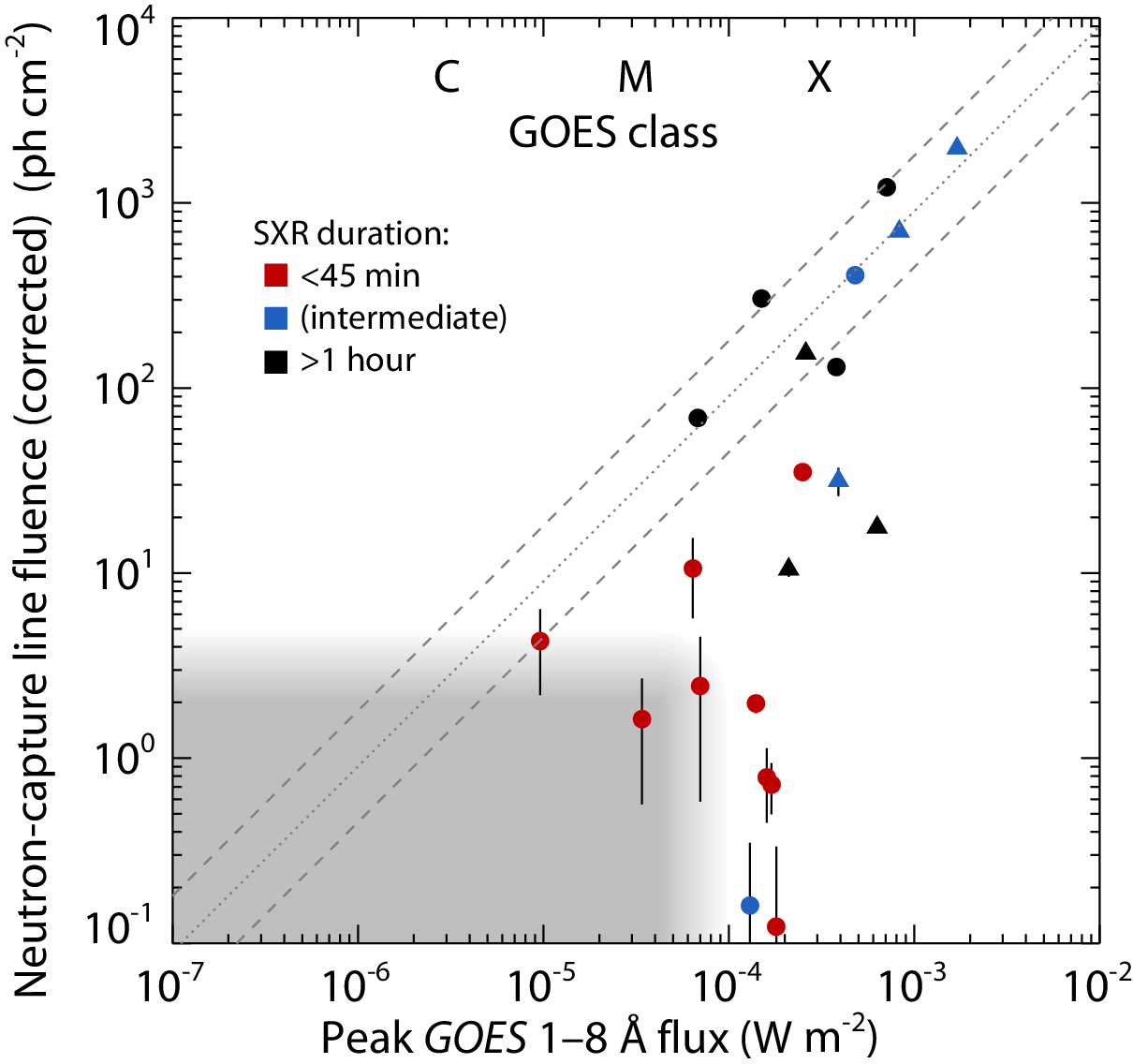}
\includegraphics[width=0.44\textwidth]{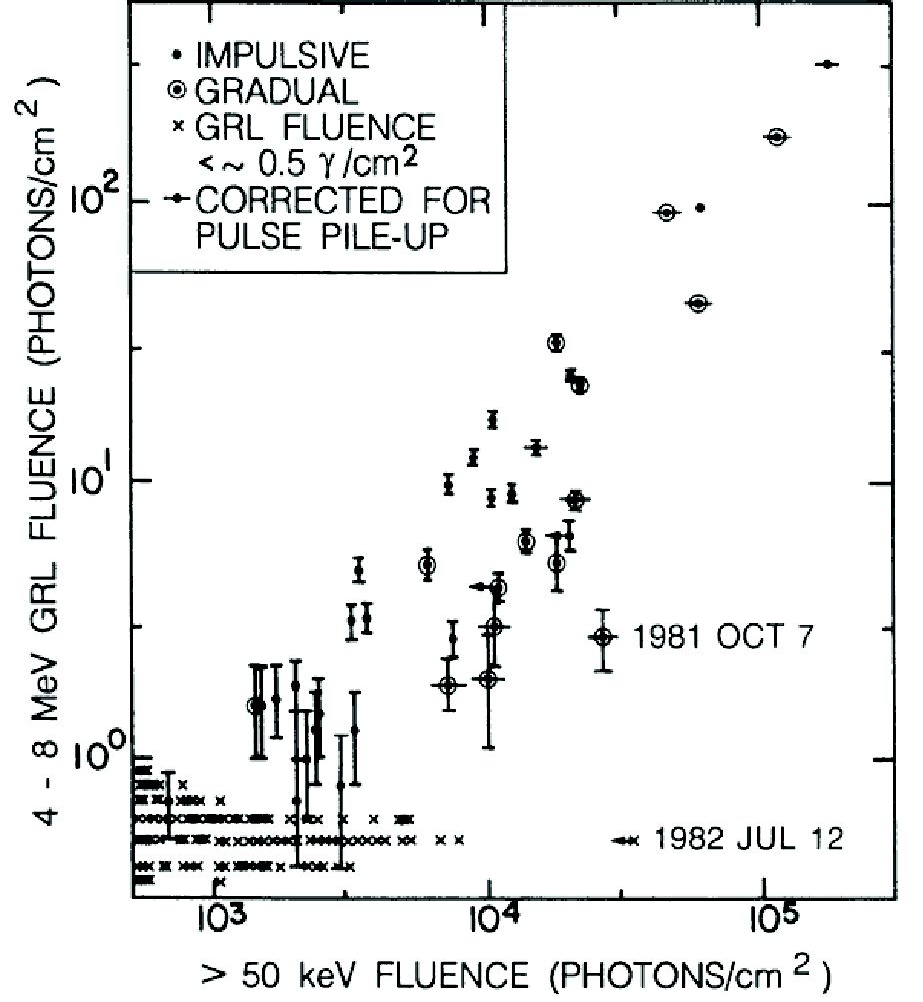}
\caption[]{\textit{Left:} the 2.223~MeV neutron-capture line fluence (indicative of the number of $\geq$30~MeV ions) plotted versus the peak \textit{\textit{GOES}} soft X-ray flux ( proportional to the non-thermal hard X-ray flux through the Neupert effect) for \textit{RHESSI} flares. 
The circles (triangles) represent flares with complete (incomplete) coverage. 
The dotted and dashed line illustrate the direct proportionality observed for the flares with the largest line fluences. The shaded area has not been systematically searched \protect\citep[from][]{2009ApJ...698L.152S}.
	 \textit{Right:} the 4-8 MeV $\gamma$-ray line plus continuum fluence (indicative of the number of $\geq$10~MeV ions) measured by the GRS instrument on SMM, plotted versus the $>$50~keV hard X-ray fluence, showing direct proportionality between $>$50 keV fluence and $\gamma$-ray line fluence for the flares with the largest line fluences  \protect\citep[from][]{1994ApJ...426..767C}.
}
\label{fig:lin_fig7}
\index{neutrons!imaging!illustration}
\end{figure}

By comparing all the $\gamma$-ray flares detected by \textit{RHESSI}, \cite{2009ApJ...698L.152S} showed that the fluence of the 2.223~MeV $\gamma$-ray line (produced by $\geq$30~MeV protons) is linearly proportional to the $>$0.3~MeV bremsstrahlung continuum fluence (produced by $>$0.3~MeV electrons), over more than three orders of magnitude in fluence, from the limit of detectability to the most intense flares \citep[Figure~2.18 of][]{Chapter4}. 
This strongly suggests that a single mechanism accelerates both $\geq$30~MeV protons and relativistic, $>$0.3~MeV electrons. 
A similarly close proportionality is observed between ion acceleration, given by either the 2.223~MeV line fluence, Figure~\ref{fig:lin_fig7} (left), or the 4-8~MeV fluence, Figure~\ref{fig:lin_fig7} (right), and tens-of-keV electron acceleration (either the fluence of 50~keV HXRs,  shown in Figure~\ref{fig:lin_fig7}, right, or the peak \textit{GOES} 1-8~ soft X-ray flux, shown in Figure~\ref{fig:lin_fig7}, left).
Here the peak soft X-ray flux is taken as a proxy for the HXR fluence -- i.e., the Neupert effect -- but only for the most powerful ion-accelerating flares, i.e., those producing more than $\sim2 \times 10^{31}$ protons $\geq$30~MeV. 
Less-powerful ion-accelerating flares, however, show a large uncorrelated excess acceleration of tens-of-keV electrons, such that even the weakest detectable ion acceleration is associated with $\sim$M-class or larger flares.\index{Neupert effect}

The observations thus are consistent with two acceleration processes: one that always accelerates both $>$30~MeV protons and $>$0.3~MeV electrons proportionally, and a second that accelerates the tens-of-keV electrons that heat the thermal flare plasma. 
These processes both occur in the flare impulsive phase, but the ion/relativistic electron acceleration may be dependent on substantial acceleration of tens-of-keV electrons (to \textit{GOES} M class). 
For the flares that are the most powerful ion accelerators, however, the fraction of energy going to tens-of-keV electron acceleration reaches a definite minimum.  

\section{Collisionless Magnetic Reconnection and Flare Energy Release}

\begin{figure}[]
\centerline{\includegraphics[width=0.9\textwidth]{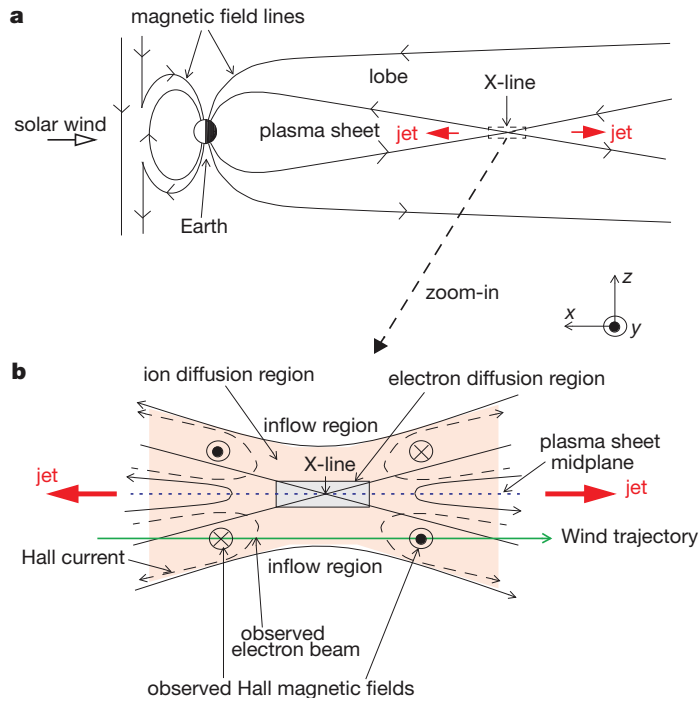}}
\caption[]{(a) The geometry of the magnetic reconnection event observed in the Earth's distant magnetotail; (b) the ion diffusion region. 
}
\label{fig:lin_fig8}
\end{figure}

The process of magnetic reconnection was first proposed by \cite{1948MNRAS.108..163G} to explain the release of magnetic energy to power a solar flare (although his discussion was in terms of currents). 
\cite{1969ARA&A...7..149S} proposed the ``neutral point theory'' for flares and \cite{1957JGR....62..509P} made the first consistent calculation of the magnetic reconnection rate for collisional resistivity.\index{resistivity}
\index{flare models!neutral-point theory} 
The Sweet-Parker reconnection rate, however, is many orders of magnitude too slow for the energy release in solar flares. 
In the last decade, significant progress has been made in the understanding of collisionless magnetic reconnection, through extensive theory and simulation work, i.e., the GEM challenge\index{GEM challenge} \citep{2001JGR...106.3737B}, \textit{in situ} space measurements, and laboratory studies.\index{magnetic reconnection!in situ@\textit{in situ} observations} 
In the collisionless case, the ions, with their large gyrodiameters, decouple from the magnetic field first, forming an ion diffusion region (Figure~\ref{fig:lin_fig8}a), while the electrons, with small gyrodiameters, are still tied to the magnetic field until they get much closer, eventually forming an electron diffusion region where they decouple from the field; the different ion and electron motions generate a current that produces a characteristic quadrupolar Hall magnetic field (Figure~\ref{fig:lin_fig8}b).\index{magnetic structures!Hall field}
The Hall reconnection rate from theory and simulations \citep{2001JGR...106.3737B} is expected to be of order 0.1~$v_{\rm A}$, compared to $\sim$10$^{-7} v_{\rm A}$ for Sweet-Parker reconnection\index{reconnection!Sweet-Parker} under solar coronal conditions \citep{2006ApJ...644L.145C}.\index{simulations!Hall MHD}\index{Alfv{\' e}n speed!and reconnection}

\begin{figure}[]
\centerline{\includegraphics[width=0.8\textwidth]{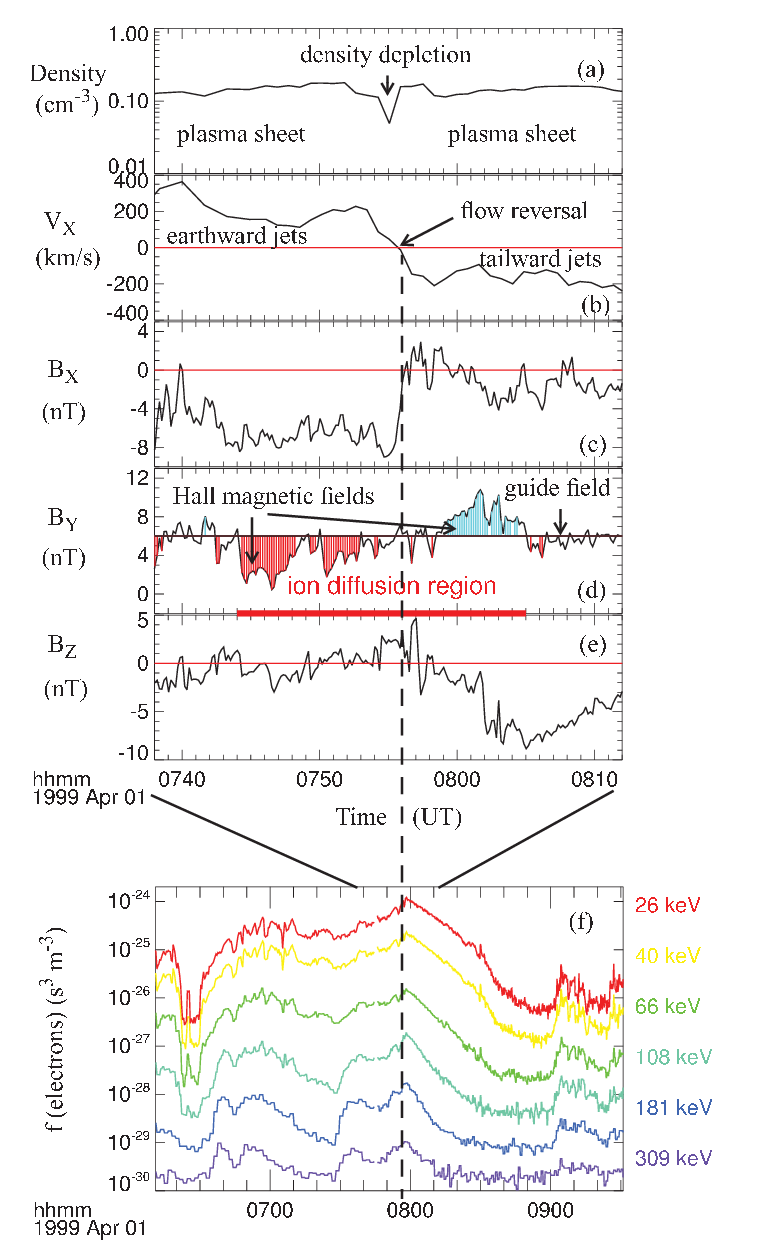}}
\caption[]{Observations by the \textit{WIND} spacecraft in the ion diffusion region (indicated by red bar in panel d) of a magnetic reconnection event \protect\citep[from][]{2001Natur.412..414O}, identified from the large 'out-of-plane' Hall magnetic fields (B$_{\rm y}$ component, panel d) surrounding the flow (proton V$_x$ in panel b) reversal. Panels c, d, \& e show the three components of the magnetic field. 
Bipolar B$_y$ field variations with polarities consistent with the Hall magnetic field pattern (see Figure~\ref{fig:lin_fig8}) were detected as the spacecraft crossed from Earthward side to the tailward side of the flow reversal region. Panel f shows that the phase space densities for 20 to $\sim$300~keV electrons peak in the ion diffusion region and decrease monotonically away from that region, with their spectrum softening with distance away. 
}
\index{reconnection!Hall!illustration}
\index{satellites!WIND@\textit{WIND}!illustration}
\label{fig:lin_fig9}
\end{figure}

Direct evidence for Hall reconnection occurring in nature was first obtained from \textit{in situ} plasma and field observations from the \textit{WIND} spacecraft\index{satellites!WIND@\textit{WIND}} (see Figure~\ref{fig:lin_fig9}) in the distant magnetotail ($\sim$60~R$_e$) of the Earth \citep{2001Natur.412..414O}\index{reconnection!Hall!magnetotail observations}.\index{magnetic structures!magnetotail}\index{magnetic reconnection!in situ@\textit{in situ} observations}
The Earth's magnetosphere/magnetotail (Figure~\ref{fig:lin_fig8}a) is similar in many respects to the standard picture of a solar flare.\index{geomagnetic tail}\index{magnetosphere}
The inner magnetosphere has closed, nearly dipolar, magnetic loops, about the same size as the loops for a large solar flare. 
The solar wind drags the outer field lines of the magnetosphere into a magnetic tail where the field is roughly oppositely directed in the north and south tail lobes, with a current sheet in between. Magnetic energy is stored in this tail, whose field strength is much larger than that of a potential dipole field.\index{current sheets!geomagnetic tail}
Transient magnetic reconnection occurring in the magnetotail leads to the release of the stored energy to produce magnetospheric substorms \citep{2008Sci...321..931A}\index{flare models!analogy with geomagnetic substorm}.  
Figure~\ref{fig:lin_fig9}, panel b, shows that the \textit{WIND} spacecraft crossed from a 100-200~km~s$^{-1}$ Earthward-directed plasma jet directly to a region of $\sim$200~km~s$^{-1}$ tailward-directed jet, implying that the spacecraft remained in the reconnection layer, crossing from one outflow jet to the other, oppositely-directed outflow jet.\index{reconnection!outflow!in situ@\textit{in situ} observations}\index{jets!reconnection!geotail}
The quadrupolar out-of-plane magnetic fields that are the signature of Hall reconnection are clearly evident (Figure~\ref{fig:lin_fig9}, panel d). 

Soon afterwards, Hall reconnection\index{reconnection!Hall!magnetospheric subsolar point} was observed by the \textit{POLAR}\index{satellites!POLAR@\textit{POLAR}} spacecraft near the subsolar point of the magnetosphere \citep{2002PhRvL..89a5002M}, and the reconnection rate was measured to be $\sim$0.02~$v_{\rm A}$ by \textit{in situ} electric-field measurements.\index{magnetosphere!reconnection rate}\index{magnetic reconnection!in situ@\textit{in situ} observations} 
Since then, many other \textit{in situ} space measurements of Hall reconnection in the Earth's magnetosphere have been reported \citep[see][for a summary]{2010JGRA..11508215E}, and even detected at Mars\index{Mars}\index{reconnection!Hall!at Mars} \citep{2009JGRA..11411204H}, indicating that this process commonly occurs in the collisionless plasmas found in nature. 
Remarkably, Hall reconnection has also been observed now in several laboratory plasma experiments (see the reviews by Yamada et al., 2010 and Zweibel \& Yamada, 2009) where conditions 
\nocite{2010RvMP...82..603Y}\nocite{2009ARA&A..47..291Z}
are completely different -- density: $n \approx\ $0.1 vs. $\sim10^{14}$~cm$^{-3}$, magnetic field: $B \approx 10^{-4}$  vs.  $\sim 100$~G, temperature: $kT_e \approx\ $400 vs. a few~eV, scale size: $\sim$100-1000~km vs. a few cm.
\index{reconnection!Hall!laboratory experiments}
This suggests that Hall reconnection is a fundamental process that can occur over a very wide range of plasma parameters, presumably including conditions in the solar corona. 
Laboratory measurements\index{reconnection!laboratory experiments} have also shown the rapid increase in reconnection rate as the plasma parameters are varied from collisional to collisionless \citep{1998PhRvL..80.3256J}.

The \textit{WIND}\index{satellites!WIND@\textit{WIND}} spacecraft also observed intense fluxes of electrons up to $\sim$300~keV energy (Figure~\ref{fig:lin_fig9}f) that peak in the ion diffusion region \citep{2002PhRvL..89s5001O} of the distant magnetotail Hall reconnection event; the energetic $>$12~keV electron fluxes are the highest observed in the $\sim$10 hour period that \textit{WIND} was in the plasma sheet. The energetic electron fluxes decrease monotonically away from this region and their spectrum softens, strongly suggesting significant electron acceleration is occurring in the ion diffusion region. No increase in energetic, $>$30 keV ion fluxes, however, is observed. 

\cite{2006Natur.443..553D} applied their model for electron acceleration by secondary islands from reconnection to this event -- they predict a power-law spectral index of $\sim$3.7 for the accelerated electron spectrum, remarkably close to the observed index of~3.8. 
More recently, for a reconnection event in the Earth's magnetotail observed {\it in situ} by the 
4-spacecraft \textit{Cluster} mission, \cite{2008NatPh...4...19C} showed that energetic electron bursts up to many tens of~keV were closely correlated with multiple islands.\index{magnetic reconnection!in situ@\textit{in situ} observations} 
Thus, electron acceleration appears to be associated with magnetic reconnection in the Earth's magnetotail, possibly with magnetic islands;
we note, however, that  other acceleration processes involving electric fields \citep{2008JGRA..11312207E} or coalescence of magnetic islands\index{magnetic structures!islands!coalescence} \citep{2010ApJ...714..915O} also can explain these observations. 
\index{reconnection!parallel electric fields}\index{electric fields!parallel} 

\cite{2005PhRvL..95w5002C} noted that both Sweet-Parker and Hall reconnection are stable solutions for a given plasma regime.\index{magnetic reconnection!Sweet-Parker!energy storage}\index{magnetic reconnection!Hall!energy release}
The slow rate of Sweet-Parker reconnection allows energy to be stored up in the magnetic field, and then a sudden transition to fast Hall reconnection can occur spontaneously when the dissipation region becomes thin enough -- from convection of stronger magnetic fields into that region \citep{2006ApJ...644L.145C} -- thus resulting in a sudden rapid energy release. 
For the Sun, \cite{2005ApJ...630..596L}, in a study of \textit{TRACE} observations of an active region reconnecting with a new flux loop, found evidence that energy was accumulated in the corona during a $\sim$24 hour phase of slow reconnection (e.g., Sweet-Parker), followed by a phase of fast reconnection with an estimated rate of $\sim$0.05~$v_{\rm A}$, presumably due to Hall reconnection in which the energy stored was released.\index{active regions!slow reconnection}

The magnetic reconnection region itself is likely to be very small, however, compared to the length of the elongated field lines in the standard flare picture or in the Earth's magnetotail, so only a very small fraction of the stored energy is released in the reconnection process itself.  
Some magnetic energy will be converted directly to particle energy through $P = |{\bf E}|I$, where ${\bf E = v \times B}$ is the convection electric field in the inflowing plasma.\index{electric fields!convective} 
The current, $I$, can be obtained from Amp{\` e}re's law. 
This goes into energizing the plasma, in principle increasing the average particle energy by $E_{\mathrm{part}} = B^2/8\pi n$ \citep[see, e.g.,][]{1977JGR....82.2761L}.
At present we cannot predict theoretically how much of this energy goes to thermal plasma and how much goes to accelerating non-thermal particles. 
The high coronal HXR source detected by \textit{RHESSI} during the pre-impulsive phase of SOL2002-07-23T00:35 (X4.8)\index{flare (individual)!SOL2002-07-23T00:35 (X4.8)!coronal hard X-ray source}\index{flare (individual)!SOL2002-07-23T00:35 (X4.8)!pre-impulsive phase}\index{pre-impulsive phase} (and several other large flares) may be the result of energization by this initial reconnection (perhaps with some additional acceleration by the collapsing magnetic field); these observations show that both thermal (superhot) plasma and accelerated electrons are produced (Figure~\ref{fig:lin_fig6}).\index{superhot component} 
The primary effect of the magnetic-reconnection process, however, is to change the topology of the field from anti-parallel fields to highly elongated turbulence, loops or islands; then the bulk of the free energy in the magnetic field can be released as the turbulence or loops or islands relax to more potential configurations.\index{magnetic fieldf!free energy}

\section{Connecting the Sun and the Heliosphere}

\subsection{Flares and impulsive SEPs}
\index{solar energetic particles (SEPs)}

The most common particle accelerations by the Sun (up to $\sim$10$^4$ per year over the whole Sun near solar maximum) are impulsive solar $\sim$1-100~keV electron events observed in the interplanetary medium \citep[see][]{1985SoPh..100..537L,2011ApJ...727..121W}. 
The SXR burst, when present, has a duration $\leq$1~hour, hence the term ``impulsive.''
These events are accompanied by low-energy ($\sim$0.01-1~MeV/nucleon) ion emissions with large enhancements of $^3$He ($^3$He/$^4$He ratio sometimes $\geq$1), heavy (e.g., Fe) and ultra-heavy (up to $\sim$200~amu) ions, with high charge (e.g., Fe$^{+20}$) states \citep[see][for a review]{2007SSRv..130..231M}. 
These events generally have relatively low particle fluxes, extend over $\sim$30-60$^\circ$ in longitude, last for hours, and have an association ($\sim$99\%) with type-III radio bursts observed in the $\sim$10~kHz-14~MHz frequency range \citep{2011ApJ...727..121W}.\index{radio emissions!type III burst!impulsive SEP events}
The very low energies of impulsive event electrons (typically $\sim$1~keV but sometimes reaching $\sim$0.1~keV) suggests that the acceleration may be occurring high in the corona \citep{1995SSRv...71..125L}, thus these events are often referred to as ``coronal flares'' \citep{1985SoPh..100..537L}.
\textit{RHESSI} (with attenuators out) has detected very weak 3-15~keV X-ray bursts in coincidence with type-III radio bursts \citep{2008ApJ...680L.149C}, suggesting a coronal explosion. 
Associations with jets observed in EUV that occur close to a coronal hole boundary \citep[Y.][]{2006ApJ...639..495W}, and with fast ($\geq$600~km~s$^{-1}$) narrow ($\leq$20$^\circ$) CMEs 
\citep{2001ApJ...562..558K,2002ApJ...579..841H,2011ApJ...727..121W} have also been reported.\index{jets!X-ray}

The e/p ratios (defined as J$_e$[0.5~MeV]/J$_p$[10~MeV]) for impulsive events, where J$_e$ and J$_p$ are the electron and proton fluxes, respectively, are comparable to those measured for $\gamma$-ray line flares \citep{1993AdSpR..13..275R,2009ApJ...698L.152S}.
Furthermore, the detailed analysis of the $\gamma$-ray spectrum of the flare SOL1981-04-2727T09:45 (X5.5) suggests that the composition of the flare-accelerated heavy ions is enhanced, similar to the composition in impulsive SEP events  \citep{1993AdSpR..13..275R,1997ApJ...479..458R,1997ApJ...489L..99M} and thus leading to the paradigm that these are due to flare acceleration \citep{1995RvGeo..33..585R}.\index{solar energetic particles (SEPs)!impulsive events}\index{paradigms!flare acceleration of impulsive SEPs}\index{flare (individual)!SOL1981-04-27T09:45 (X5.5)!gamma-ray@$\gamma$-ray spectrum}
Only $\sim$25\% of impulsive events, however, have an associated flare reported by \textit{GOES} (although the association is much higher when compared with SXR bursts detected by \textit{RHESSI}; see Krucker et al., 2007b), and only a very few events have been detected from $\gamma$-ray line flares; \textit{RHESSI} has not detected  $\gamma$-ray line emission from the flares associated with the impulsive events\index{gamma-rays!non-detection in impulsive events}. 
\nocite{2007ApJ...663L.109K}

By assuming that the impulsive event electrons and ions at all energies travel the same path length~$L$ from the Sun to the observation site (typically near $\sim$1~AU) -- i.e., $L = v(E) (t(E) - t_0$), both the injection time $t_0$ and $L$~can be derived from the observed event onset times for particles with different velocities \citep{1974SSRv...16..189L}. 
Surprisingly, for most ($\sim$80\%) of the events, the derived injection times at energies above $\sim$20~keV are delayed by $\sim$10-30 minutes from the start of the associated type III burst 
\citep{1999ApJ...519..864K,2002ApJ...579..841H}, while $\sim$20\% of the events are prompt (no delay).
\index{particles!delayed}\index{solar energetic particles (SEPs)!delayed}
Interestingly, in the delayed events, the injection of lower-energy ($\sim$1-10~keV) electrons begins before the start of the associated type~III burst at the Sun \citep[L.][]{2006ApJ...639..495W}. 
As these electrons escape from the Sun, their velocity dispersion results in a bump-on-tail distribution that generates Langmuir waves that in turn produce the associated type~III radio burst.   
Consistent with this picture, intense Langmuir waves are detected \textit{in situ} at $\sim$1~AU only when these $\sim$1-10 keV electrons arrive \citep{1998ApJ...503..435E}\index{waves!Langmuir!in situ@\textit{in situ} detection}. 

The energy spectra of impulsive event electrons are typically double power laws with a downward break at tens of keV \citep{2009ApJ...691..806K}, very similar in shape to the non-thermal HXR spectra typically observed by \textit{RHESSI} for flares.  For prompt events, a direct comparison of \textit{RHESSI} HXR spectrum with the escaping electron spectrum above $\sim$50 keV (above the break) shows that the power-law spectral indices are linearly correlated with $\gamma_x  \approx \delta_e$ for prompt events (and uncorrelated for delayed events)\index{acceleration!delayed}.
This is clearly different from the $\gamma_x  + 1  \approx \delta_e$ that would be expected if the electrons escaping to the interplanetary medium come from the single acceleration high in the corona, while most of the accelerated electrons go down to the chromosphere to produce the HXR emission in losing all their energy to collisions (thick target). The number of electrons escaping the Sun to produce the impulsive event is typically only of order $\sim$0.1\% of the number required to produce the HXR burst. 

\textit{RHESSI} HXR imaging of the flares associated with impulsive events \citep{2010AAS...21632006G} indicates that reconnection between open field lines and closed loops
\citep[interchange reconnection -- see][]{1977ApJ...216..123H,1996AdSpR..17..197S} is involved. 
\index{reconnection!interchange}
This is consistent with a model where the electrons accelerated in this reconnection process that go upward escape to the interplanetary medium, while those going downward are trapped on closed field lines and are further accelerated as the field lines shrink \citep[e.g.][]{2004A&A...419.1159K} modifying the spectrum and increasing the number of electrons above 50~keV to produce the observed HXR emission. 
Thus, the escaping electrons (and perhaps the escaping ions) may be a sample of the particle energization in the flare magnetic reconnection process; note that the electron spectrum below $\sim$50~keV at $\sim$1~AU could also have been modified by wave-particle interactions in propagating to $\sim$1~AU \citep[e.g.,][]{2010ApJ...721..864R}.
\index{wave-particle interactions}

\subsection{Flares, CMEs, and gradual SEP ervents}
\index{solar energetic particles (SEPs)!gradual}

The most powerful ion accelerators (sometimes ions reach 10~to $\sim$100~GeV, the most energetic particles accelerated in the solar system) are large gradual (associated flare SXR burst $\geq$1~hour duration) SEP events, that occur on average about once per month near solar maximum, and are dominated by intense fluxes of $>$10~MeV protons with e/p ratios about two orders of magnitude lower than the ratios for $\gamma$-ray line flares \citep{1992ApJ...391..370K,2007ApJ...658.1349C,2009ApJ...698L.152S}.
\index{accelerated particles!electron-to-proton ratio}
These events have generally normal coronal abundances and ionization states \cite[see][for a revew]{2009IAUS..257..401C}, extend over $\sim$100-180$^\circ$ in solar longitude, last for days, and are closely associated with fast CMEs and with type~II radio bursts, suggesting that the SEPs are accelerated by shocks driven by fast CMEs and not by flares \citep{1995RvGeo..33..585R}.\index{ionization state!coronal signature}\index{abundances!SEPs}
\index{coronal mass ejections (CMEs)!abundances}\index{abundances!CMEs}\index{space weather} 
Fast CMEs are essentially always accompanied by simultaneous large flares; these solar eruptive events are the most powerful explosions and most energetic particle accelerators in the solar system, and they produce the most extreme space weather. 
The observed delays in gradual SEP event onset times \citep{1994ApJ...428..837K}, and the agreement of composition and temporal variations of the particle fluxes \citep{2006ApJ...646.1319T} with theoretical models indicate that diffusive shock acceleration (the same process that is believed to accelerate galactic cosmic rays\index{cosmic rays!supernova shock waves} in supernovae\index{supernovae} shock waves) at altitudes of $\sim$2-40 
~R$_\odot$ is responsible for gradual SEPs.\index{acceleration!diffusive shock acceleration}\index{acceleration!shock}\index{shocks!supernova} 
For large, fast CMEs, the number of accelerated ions is sometimes significantly larger than the number required to produce the $\gamma$-ray line emission in the associated flare, and the total energy in the SEPs is typically of order $\sim$10\% of the total CME energy that is dominated by the kinetic energy of the fast ejecta \citep[see Figure~6.3 of][]{Chapter2}.\index{accelerated particles!energy content of}
Such a high efficiency is required for supernovae shock acceleration to produce galactic cosmic rays. 

Recent studies show that acceleration profiles of fast CMEs are synchronous with, and closely resemble, the flare energy release as measured by the \textit{RHESSI} hard X-ray flux, \citep[Figure~5.2 of][]{Chapter2}; this is consistent with the standard model for large solar eruptions of magnetic reconnection occurring in a current sheet behind the CME.\index{standard model}\index{current sheets!and CMEs}
\textit{RHESSI} has detected a large ($\sim1.5 \times 10^5$~km diameter) high coronal non-thermal HXR source that is expanding (at $\sim$400~km~s$^{-1}$) and moving outward at $\sim$750~km~s$^{-1}$ behind a very fast ($\sim$2300~km~s$^{-1}$) CME whose associated flare was $\sim$40$^\circ$ behind the limb \citep{2007ApJ...669L..49K}. 
The HXRs are emitted by non-thermal, $>$10~keV electrons (constituting $\sim$10\% of the total electron density) trapped in closed magnetic structures related to the CME. 
Furthermore, \textit{RHESSI} has detected non-thermal HXR emission from every fast 
($>$1500~km~s$^{-1}$) ``backside'' CME where the associated flare site is 20$^\circ$ to 50$^\circ$ behind the limb.    
\index{coronal mass ejections (CMEs)!backside}

\cite{1995ApJ...453..973K} found that flares with HXR burst spectra that evolved in time from soft to hard to harder (SHH, as opposed to the standard SHS behavior of most flares) are closely associated with high-energy SEP\index{solar energetic particles (SEPs)!HXR signature} events observed in interplanetary space.
\index{soft-hard-soft}\index{soft-hard-harder}\index{hard X-rays!soft-hard-soft}\index{hard X-rays!soft-hard-harder}
In a statistical study of all \textit{RHESSI} flares, \cite{2009ApJ...707.1588G} drew the same conclusion -- they found that all \textit{RHESSI} flares associated with an SEP event show SHH behavior, and none of the flares with SHS (normal) behavior are associated with an SEP event\index{coronal mass ejections (CMEs)!and soft-hard-harder pattern}. \textit{RHESSI} images show that the HXRs originate from footpoints during times with SHH behavior
\citep{2008ApJ...673.1169S,2008ApJ...683.1180G}, and there is no abrupt change in footpoint motion at the onset of the hardening, indicating the same acceleration process taking place in the preceding impulsive phase appears to continue into the hardening phase. 
Furthermore, for a few flares that are well connected (W30-90), the energetic ion spectra inferred from the \textit{RHESSI} $\gamma$-ray line observations appear similar to the spectra of SEPs observed near the Earth, suggesting the possibility that flare acceleration may contribute to gradual event SEPs on well-connected field lines. 
Thus, in large solar eruptive events there is evidence for a close physical connection between flares, fast CMEs, and SEP acceleration.

\section{Summary}

\textit{RHESSI's} HXR and $\gamma$-ray line imaging and previous observations have provided strong evidence that flare electron and ion acceleration is related to magnetic reconnection, and \textit{RHESSI's} spectroscopy and imaging have confirmed that in many flares most of the energy released is contained in the accelerated electrons and ions. 
Now a reconnection process fast enough for the release of energy in a flare -- collisionless Hall reconnection -- has been identified through theory and simulations, and directly observed in space and laboratory measurements to operate over a very wide range of plasma parameters. 
Presumably, magnetic energy is being stored in the corona by the motions of footpoints (or the emergence of new bipole from below the photosphere), when conditions are such that slow Sweet-Parker collisional reconnection applies.  
Then the flare begins with a spontaneous transition from slow Sweet-Parker collisional reconnection to fast collisionless Hall reconnection high in the corona, when the incoming magnetic fields thin the current sheet between regions of different magnetic direction sufficiently.\index{current sheets!and Hall reconnection}
This fast reconnection will start in one small region, but based on simulations, reconnection will rapidly develop at multiple different sites. 
The high coronal HXR sources detected by \textit{RHESSI} during the pre-impulsive phase of large flares (such as SOL2002-07-23)\index{flare (individual)!SOL2002-07-23T00:35 (X4.8)!reconnection}
 may be the result of energization by this initial reconnection (perhaps with some additional acceleration by the collapsing magnetic field\index{collapsing magnetic trap}); both thermal (superhot\index{superhot component}) plasma and accelerated electrons are produced. 
 When the reconnection occurs with open field lines (interchange reconnection), this direct energization may provide the $\sim$1-10~keV electrons in impulsive events observed in the interplanetary medium. 
The upward reconnection jet may result in a fast narrow CME that then accelerates ions and further accelerates electrons higher up in the corona, leading to the observed delays \citep[e.g.,][]{2011ApJ...727..121W}. 

The impulsive phase of the flare is characterized by the most intense and energetic non-thermal HXR continuum and $\gamma$-ray line emissions, mostly in the footpoints, implying that the strongest energy release and particle acceleration is occurring then, and that the energetic particles are precipitating into the chromosphere and photosphere. 
The high coronal HXR source observed in the Masuda flare\index{flare (individual)!SOL1992-01-13T17:25 (M2.0)!coronal hard X-ray source} above the soft X-ray looptops provided the first indication that the main flare energy release/particle acceleration occurred there. 
The occulted flare SOL2007-12-31T01:11 (C8.3) had a similar HXR coronal source above the looptops, but here \textit{RHESSI's} high spectral resolution was able to show that this source was filled with energetic electrons with a power-law spectrum (with no thermal plasma detectable above background), indicating that most of the electrons in the source had been accelerated.\index{occulted sources}
\index{flare (individual)!SOL2007-12-31T01:11 (C8.3)!above-the-loop-top source}
\index{electrons!dominant tail population}\index{electrons!distribution function!purely nonthermal}
The  bulk of the electrons being convected into this region at a speed of $\sim$0.1~$v_{\rm A}$  (as expected for inflows to a magnetic reconnection region) need to be accelerated to provide the rate of electrons per second required to produce the observed footpoint HXR emission. 

Simultaneous multi-frequency microwave imaging showed that, in this source, the electron power-law spectrum extended to $\sim$1~MeV and the magnetic field is $\sim$30-50~G.\index{frequency!microwave peak}
The energy density of the accelerated electrons above $\sim$16~keV is then about equal to the magnetic energy density, i.e. $\beta \sim 1$. 
The upper limits on the X-ray emission from this region before the impulsive phase indicate that 
$\beta \sim 0.01$,  implying a mechanism -- such as that proposed by \cite{2006Natur.443..553D} -- that very efficiently converts the stored magnetic energy, presumably in the form of elongated magnetic fields, to bulk acceleration of electrons.\index{distribution function!purely nonthermal}
It is important for any theory of the acceleration of energetic particles in flares to show how they are related to magnetic reconnection, and how they can put the bulk of the energy released in a flare into energetic particles. 
In particular, acceleration mechanisms that rely on an intermediary such as waves, turbulence, or shocks to accelerate the particles, must show how a high overall efficiency -- the efficiency for generating the intermediary, multiplied by the efficiency of the intermediary in accelerating the particles -- can be achieved.

At present, we do not understand what triggers the sudden onset of the impulsive phase, or how such large numbers of electrons propagate down to the chromosphere. 
A possible clue is the abrupt change from downward motion of the coronal HXR source during the pre-impulsive phase of SOL2002-07-23\index{flare (individual)!SOL2002-07-23T00:35 (X4.8)!source motions}, to upward motion at the onset of the impulsive phase. 
The downward motion suggests collapsing or shortening of magnetic field lines, which should result in further energization of particles in the source\index{coronal sources!contraction}. 
Perhaps when the source gets low enough the bulk energization of electrons is triggered, and the accelerated electrons are then able to propagate to the footpoints to produce the observed HXR emission\index{acceleration!bulk energization}.  

A single mechanism -- also related to magnetic reconnection and occurring in the flare region during the impulsive phase -- appears to accelerate both $>$30~MeV ions and relativistic, $>$0.3~MeV electrons proportionately. 
This mechanism, however, results in the spatial displacement of the accelerated ions from electrons; perhaps electric fields are involved. 
Significant acceleration of tens-of-keV electrons (enough to produce soft X-ray emission of \textit{GOES} M-class or larger) appears to be a prerequisite to this acceleration, but when very large numbers of ions/relativistic electrons are accelerated, the acceleration of tens-of-keV electrons becomes proportional as well. 
At present, none of the current particle acceleration models \citep[see][]{Chapter8} provide a compelling explanation for these correlations between accelerated electrons and ions, and for their differences in spatial morphology. 

Finally, the \textit{RHESSI} observations show that large flares appeared to be closely connected to their associated fast CMEs and the acceleration of SEPs. 
At present we do not understand the physical mechanisms underlying this connection. 

\section{Future Prospects}

Much higher sensitivity HXR imaging spectroscopy\index{imaging spectroscopy!improvements} is clearly needed to study the temporal, spatial, and spectral evolution of the high coronal sources in typical flares, rather than just the ones for which that source is anomalously bright.
The \textit{RHESSI} observations show that this key region is where the energy release/electron acceleration appears to be happening.
High sensitivity will also allow the detection of  the HXR emission from the accelerated electrons as they travel down the legs of the loops to the chromosphere; such measurements will help to understand the propagation of these large numbers of electrons through the atmosphere. 
These measurements should also have much larger dynamic range than \textit{RHESSI}, so the coronal sources can be observed simultaneously with the very bright footpoint sources. 
In the past decade, focusing optics have been developed for HXRs up to $\sim$80~keV, and now can provide angular resolutions of $\sim$7$''$, fine enough for solar measurements. 
A Focusing Optics hard X-ray Spectrometer Imager (FOXSI) instrument is presently being developed for a rocket flight in late 2011 \citep{2009SPIE.7437E...4K}.\index{FOXSI}
A \textit{FOXSI}-like instrument can provide much higher sensitivity and dynamic range than \textit{RHESSI}. 

Much more sensitive solar $\gamma$-ray line imaging with higher spatial resolution is required to follow the temporal, spatial, and spectral evolution of the ion footpoints, to make progress in understanding the acceleration of ions in solar flares. 
A Gamma-Ray Imaging Polarimeter for Solar flares (GRIPS)\index{GRIPS} instrument that utilizes cooled germanium detectors with 3-D spatial resolution of 0.5~mm, together with a single rotating grid of novel design mounted $\sim$10~m away to provide the imaging spectroscopy, is presently being developed for a first balloon flight in 2012 \citep{2009SPD....40.1810S}. 

Now that $\sim$1.8-5 MeV energetic neutral atoms\index{energetic neutral atoms} (ENAs) have been detected from an SEP event
\citep{2009ApJ...693L..11M} for the first time, it appears possible to image SEPs above $\sim$2~R$_\odot$ -- where they are presumably being accelerated by the fast CME shock wave -- via the ENAs they produce through charge exchange.\index{reactions!charge-exchange}\index{shocks!and ENAs}
ENAs cannot be focused to form an image, but the \textit{RHESSI} technique of Fourier-transform imaging using modulation methods can be applied. 

The above measurements of the accelerated particles need to be combined with imaging measurements of the magnetic field plus plasma parameters -- density, temperature, and flows -- in the corona where the energy release/particle acceleration is taking place. 
The plasma parameters can be obtained with a combination of UV/EUV/soft X-ray imaging spectroscopy,\index{imaging spectroscopy!multi-band} while new techniques have been developed to measure the coronal magnetic fields through radio (e.g., FASR) and optical (CoSMO and ATST) imaging spectroscopy. 
Thus, the next solar maximum (estimated $\sim$2023) would be an ideal time to make the great leap forward in understanding the fundamental physics of transient energy release and efficient particle acceleration in large solar eruptions (and therefore in cosmic magnetized plasmas) especially since complementary measurements of the solar energetic particles escaping to space will likely be provided by the upcoming NASA \textit{Solar Probe Plus} mission and ESA/NASA \textit{Solar Orbiter} mission (which also provides coronagraphic and other imaging measurements) are planned to be going close to the Sun at the time of the next solar maximum. 
Planning for a \textit{Solar Eruptive Events (SEE)} mission that provides the above-mentioned measurements is presently under way as part of the 2013 Heliophysics Decadal Survey. 

\begin{acknowledgements}
I'm pleased to acknowledge illuminating discussions and assistance from S. Krucker, H. Hudson, G. Emslie, A. Caspi, A. Shih, M. {\O}ieroset, L. Wang, T. Phan, J. Drake, J. Egedal, and L. Glesener. This research was supported in part by NASA contract 5-98033 and by the WCU grant no. R31-10016 funded by the Korean Ministry of Education, Sciences, and Technology. 
\end{acknowledgements}

\bibliographystyle{ssrv}

\bibliography{ch9}

\printindex

\end{document}